\documentclass[%
 aip,
 amsmath,amssymb,
 reprint,%
]{revtex4-1}

\usepackage{graphicx}
\usepackage{dcolumn}
\usepackage{bm}

\usepackage[utf8]{inputenc}
\usepackage[T1]{fontenc}
\usepackage{mathptmx}
\usepackage{etoolbox}

\usepackage{gnuplottex}
\usepackage{mathtools}
\usepackage{float}
\usepackage{array}
\usepackage[english]{babel}
\usepackage[utf8]{inputenc}
\usepackage{amsmath}
\usepackage[table,dvipsnames]{xcolor}
\usepackage{amsfonts}
\usepackage{graphicx}
\graphicspath{ {./Figures/} }
\usepackage{dsfont}
\usepackage[colorlinks=true, linkcolor=blue, citecolor=blue]{hyperref}
\usepackage{algorithm}
\usepackage[noend]{algpseudocode}
\usepackage{bbold}
\usepackage{mathtools}
\usepackage{amssymb}
\usepackage{bm}
\usepackage{bbm}
\usepackage{tabularx, booktabs}
\usepackage{xspace}
\usepackage{listings}
\usepackage{tikz}
\usepackage{bbding}
\usepackage{physics}
\usepackage{placeins} 
\usepackage{standalone}
\usetikzlibrary{arrows.meta}
\usetikzlibrary{shapes}
\usepackage{ulem}

\usepackage{multirow}

\AtBeginEnvironment{bmatrix}{\setlength{\arraycolsep}{2.5pt}}

\makeatletter
\def\mathcolor#1#{\@mathcolor{#1}}
\def\@mathcolor#1#2#3{%
  \protect\leavevmode
  \begingroup
    \color#1{#2}#3%
  \endgroup
}
\makeatother

\newcommand{\bgam}{\bm{\gamma}}
\newcommand{\bgamt}{\bm{\tilde\gamma}}

\newcommand{\bw}{\mathbf{w}}
\newcommand{\bP}{\mathbf{P}}
\newcommand{\bQ}{\mathbf{Q}}

\newcommand{\be}{\begin{eqnarray}}
\newcommand{\ee}{\end{eqnarray}}
\newcommand{\myket}[1]{\left\vert #1\right\rangle}
\newcommand{\mykettight}[1]{\vert #1\rangle}
\newcommand{\mybra}[1]{\left\langle #1\right\vert}

\newcommand{\brakettight}[3]{\langle #1 \vert #2 \vert #3 \rangle}
\newcommand{\mybraket}[3]{\left\langle #1 \middle\vert #2 \middle\vert #3 \right\rangle}

\newcommand{\gamt}{\tilde\gamma}

\newcommand{\cdop}{\hat{c}^{\dagger}}
\newcommand{\cop}{\hat{c}}

\newcommand{\rev}[1]{{#1}}

\makeatletter
\def\@email#1#2{%
 \endgroup
 \patchcmd{\titleblock@produce}
  {\frontmatter@RRAPformat}
  {\frontmatter@RRAPformat{\produce@RRAP{*#1\href{mailto:#2}{#2}}}\frontmatter@RRAPformat}
  {}{}
}%
\makeatother

\begin{document}

\preprint{AIP/123-QED}





\title[]{ 
Fragment quantum embedding using the Householder transformation: a multi-state extension based on ensembles
}

\author{Filip Cernatic$^*$} 
\email{filip.cernatic@gmail.com}
\affiliation{Laboratoire de Chimie Quantique, Institut de Chimie,
CNRS/Université de Strasbourg, 4 rue Blaise Pascal, 67000 Strasbourg, France}

\author{Emmanuel Fromager} 
\affiliation{Laboratoire de Chimie Quantique, Institut de Chimie,
CNRS/Université de Strasbourg, 4 rue Blaise Pascal, 67000 Strasbourg, France}

\author{Saad Yalouz}
\affiliation{Laboratoire de Chimie Quantique, Institut de Chimie,
CNRS/Université de Strasbourg, 4 rue Blaise Pascal, 67000 Strasbourg, France}

\date{\today}

\begin{abstract}
In recent works by Yalouz et al. (J. Chem. Phys. 157, 214112, 2022) and Sekaran et al. (Phys. Rev. B 104, 035121, 2021; Computation 10, 45, 2022), Density Matrix Embedding Theory (DMET) has been reformulated through the use of the Householder transformation as a novel tool to embed a fragment within extended systems. The transformation was applied to a reference non-interacting one-electron reduced density matrix to construct fragments' bath orbitals, which are crucial for subsequent ground state calculations. In the present work, we expand upon these previous developments and extend the utilization of the Householder transformation to the description of multiple electronic states, including ground and excited states. Based on an ensemble noninteracting density matrix, we demonstrate the feasibility of achieving exact fragment embedding through successive Householder transformations, resulting in a larger set of bath orbitals. We analytically prove that the number of additional bath orbitals scales directly with the number of fractionally occupied natural orbitals in the reference ensemble density matrix. A connection with the regular DMET bath construction is also made. Then, we illustrate the use of this ensemble embedding tool in single-shot DMET calculations to describe both ground and first excited states in a Hubbard lattice model and an \textit{ab initio} hydrogen system. Lastly, we discuss avenues for enhancing ensemble embedding through self-consistency and explore potential future directions.

\end{abstract}

\maketitle


\section{Introduction}\label{sec:Introduction}

Density Matrix Embedding Theory (DMET) is a computational method initially introduced to investigate the ground state of strongly correlated many-electron systems within condensed matter physics and quantum chemistry~\cite{knizia2012density, knizia2013density, wouters2016practical, wouters2017five}. 
From a practical point of view, the ambition with DMET is to simplify and replace large-scale calculations with a series of more manageable smaller-sized problems. Starting with an extended system (e.g., a large molecule or lattice), the procedure typically begins with a partitioning of the target system into smaller local fragments. 
Once this partitioning is completed, the objective is then to assess how each local fragment interacts with its environment to recover part of the electronic correlations. 
To proceed, one usually chooses to switch to a more convenient representation of the system, the so-called "fragment+bath" picture, which drastically simplifies our vision of the full problem.
The "bath" is an effective small-size system containing a small number of orbitals (non-overlapping with the fragment) whose goal is to mimic the fragment’s surrounding.
The union of the fragment's and bath's orbital spaces forms a so-called "\textit{cluster}" amenable to high-level theoretical methods (e.g., Configuration Interaction) due to its reduced size compared to the whole system.
Resolving the reduced-in-size Schrödinger equations attached to each cluster provides insights on the ground state properties of the system at a lower numerical cost. 
Over the past decade, numerous examples have demonstrated the efficiency of the DMET procedure in yielding accurate ground state results for various types of systems. Illustrative cases of applications encompass lattice models and periodic systems~\cite{zheng2016ground, bulik2014density, JCTC20_Chan_ab-initio_DMET, pham2019periodic, wu2019projected, sekaran2021householder, sekaran2022local,cui2023abinitioquantummanybody,Booth24_PRL_Rigorous_Screened_Interactions}, large-sized molecules and reactions~\cite{li2023multi, wouters2016practical, pham2018can}, and even extensions to hybrid fermion-boson systems~\cite{reinhard2019density, sandhoefer2016density} to cite but a few (see also the reviews~\cite{sun2016quantum,wouters2017five} and the references within). At the theory level, DMET has been recently reviewed from a mathematical perspective ~\cite{cances2023mathematicalinsightsdensitymatrix}, with a particular focus on the convergence of its self-consistency loop. Connections with alternative embedding techniques, such as the Ghost Gutzwiller approach~\cite{Lanata23_Derivation_ghost_Gutzwiller,Mejuto-Zaera24_Faraday_ghost_Gutzwiller} or the exact factorization~\cite{lacombe2020embedding,requist2021fock-space}, have also been explored.

 It is noteworthy that most of the DMET developments present in the literature mainly revolve around ground state applications. Conversely, only a few works have attempted to extend the DMET framework to the field of excited states.
In this context, one can mention the work of Tran et al., who proposed a state-specific embedding method that combines DMET with SCF metadynamics within quantum chemistry contexts~\cite{tran2019using}. In this approach, an initial set of molecular orbitals is optimized for a specific excited state (in a mean-field way) to guide the future fragment embedding. Similarly, techniques called 'bootstrap embedding' have also been utilized for state-specific descriptions of excited states in molecules~\cite{ye2021accurate}.
In another case, to explore excited states of crystalline point defects in a state-specific way, Mitra et al. proposed a periodic DMET calculation based on restricted and open-shell Hartree-Fock calculations to identify bath orbitals~\cite{mitra2021excited}. Expanding beyond molecular systems to structures like Hubbard lattices, developments have also been made to extend DMET to response wavefunctions, thereby enabling the computation of spectral functions~\cite{chen2014intermediate,booth2015spectral}.
Finally, in opening up the question of ensemble embedding, a finite-temperature version of DMET has already been proposed~\cite{sun2020finite}. In this approach, an approximate bath was introduced to reproduce the mean-field finite-temperature 1-electron Reduced Density Matrix (1-RDM) of an ensemble of electronic states. In the present work, we will focus on a different type of ensemble that is referred to as Gross-Oliveira-Kohn (GOK) ensemble~\cite{gross1988rayleigh}. In practical calculations, the latter usually contains a few (low-lying) states to which individual ensemble weights are arbitrarily assigned, unlike in a thermal ensemble.

 
As depicted here, the development of DMET-like methods for excited states remains an emerging field of research. In this context, a predominant focus has been placed on state-specific methodologies~\cite{tran2019using,ye2021accurate,mitra2021excited,chen2014intermediate,booth2015spectral}, while only a limited number of studies have explored multi-state embedding~\cite{sun2020finite}. In the present paper, we aim to fill this gap by introducing a strategy for automatically constructing optimal fragment bath orbitals adapted to multiple electronic states concurrently.
Our strategy leverages the so-called Householder transformation~\cite{householder_unitary_1958,householder1958generated,martin1968householder} to generate exact fragment embedding for multi-state calculations at the non-interacting level. 
This approach contrasts with the original version of DMET~\cite{knizia2012density,knizia2013density}, where the primary embedding transformation proposed to create the "fragment+bath" subspace was the Singular Value Decomposition (SVD). 
Directly inspired by the Schmidt decomposition~\cite{nielsen2001quantum} (an important tool in quantum information theory~\cite{nielsen2001quantum}), the SVD represents a maximally disentangling transformation that ensures the identification of optimal bath orbitals, mimicking the environment of a specified fragment. As an alternative to SVD, the Householder transformation~\cite{householder_unitary_1958} was recently highlighted by some of the present authors as a practically simpler (but theoretically equivalent~\cite{sekaran2023unified}) embedding transformation~\cite{yalouz2022quantum,sekaran2021householder,sekaran2022local}.
When applied to an idempotent (non-interacting) 1-RDM, the Householder transformation was shown to optimally block-diagonalize the matrix into two independent blocks: one for the cluster and another for the rest of the system. This approach is straightforward as it automatically yields the optimal bath orbitals of a given fragment without further analysis of the 1-RDM, contrasting with SVD, which requires more manipulations (see Section~2.2 of the practical guide on DMET~\cite{wouters2016practical}; see also Ref.~\citenum{sekaran2023unified}). Despite this practical difference, a recent work has demonstrated that the resulting bath subspace produced by the Householder transformation is identical to the one obtained with SVD~\cite{sekaran2023unified}, and thus optimal in the sense of the Schmidt decomposition.

In the present work, we expand upon these previous developments and extend the utilization of the Householder transformation to the description of multiple electronic states.
Based on an ensemble noninteracting density matrix, we will demonstrate the feasibility of achieving exact fragment embedding through successive Householder transformations, resulting in a larger set of bath orbitals.
We will show that the number of additional bath orbitals associated with the fragment scales directly with the number of fractionally occupied natural orbitals in the reference ensemble density matrix. 
The method will be thoroughly detailed theoretically and numerically, and also illustrated on both model and \textit{ab initio} Hamiltonians.  
To facilitate the reproducibility of all our results, we use the open access python package QuantNBody~\cite{yalouz2022quantnbody} dedicated to the manipulation of quantum many-body systems and recently proposed by one of the authors (SY). 
This package allows a systematic construction and diagonalization of the effective
embedding Hamiltonians associated to the cluster space. 
It also includes our own numerical implementation of the Householder transformation as a ready-to-use function.  
 
  The manuscript is organized as follows. In Section~\ref{sec:Theor}, we discuss all the theoretical aspects of the Householder-based (single-orbital) fragment ensemble embedding, beginning with a brief overview of the ground state approach and switching then to the multi-state approach. In Section~\ref{eq:comparison_with_DMET}, we rationalize our main findings by discussing the extension of the regular DMET quantum bath construction to ensembles (with no restrictions in the size of the to-be-embedded fragment).     
  In Section~\ref{sec:Applications}, we present an application of our embedding strategy for ground and first excited state calculations in two illustrative examples, namely, a Hubbard ring and an \textit{ab initio} system of hydrogen atoms. Finally, we conclude in Section~\ref{sec:Conclusion} with a summary of the main theoretical ideas and key observations, and we point out possible routes for future improvements.

\section{Householder-Based Fragment Embedding}\label{sec:Theor}

Starting from the 1-electron Reduced Density Matrix (1-RDM),  we will detail in this section how the Householder transformation can be employed to build the bath orbitals for a given system's fragment.
We will first discuss the simplest scenario where a fragment is embedded to describe a single state (i.e. the ground state).
Then, we will discuss the changes and difficulties arising when switching to an ensemble of states, and show that the Householder transformation can still be used in this framework to implement an exact fragment embedding.
In the following, we will assume a generic many-electron system composed of $N$ electrons and $L$ spatial orbitals in total.

\subsection{Exact fragment embedding for ground state at the non-interacting level} \label{sec:GS}

In regular ground state DMET, the central object employed to generate the embedding of a given local fragment is the 1-RDM denoted by $\boldsymbol{\gamma}$. Usually expressed in a local spin orbital basis, this matrix is defined by:
\begin{equation}\label{eq:1-RDMdef}
\gamma_{pq}^{\Phi_0} = \brakettight{\Phi_0}{\cdop_{p}\cop_{q}}{\Phi_0},
\end{equation}
where $\cdop_{p}$ and $\cop_{p}$ represent local spin-orbital creation and annihilation operators, respectively. In this definition, $\ket{\Phi_0}$ typically represents a single Slater determinant obtained from a non-interacting mean-field calculation, such as Hartree-Fock. This state serves as an approximation of the exact ground state $\ket{\Psi_0}$ of the interacting system, which is challenging to access in practice. Consequently, the 1-RDM given in Eq.~(\ref{eq:1-RDMdef}) is idempotent, satisfying $\gamma^{\Phi_0} = (\gamma^{\Phi_0})^2$ (this property is crucial for the subsequent analysis). The expression of the 1-RDM in a local basis (as given in Eq.~(\ref{eq:1-RDMdef})) is necessary to define
reference local fragments, consisting of $L_F<2L$ spin-orbitals each, that are to be embedded into the extended system. 
In the
simplest scenario where an
individual fragment contains $L_F=2$ spin-orbitals sharing the same local spatial orbital (single-orbital fragment), the so-called Householder transformation \cite{householder_unitary_1958}
affords a straightforward
construction of its associated bath spin-orbitals.

\begin{figure}[t!]
\centering
\includegraphics[width=0.49\textwidth]{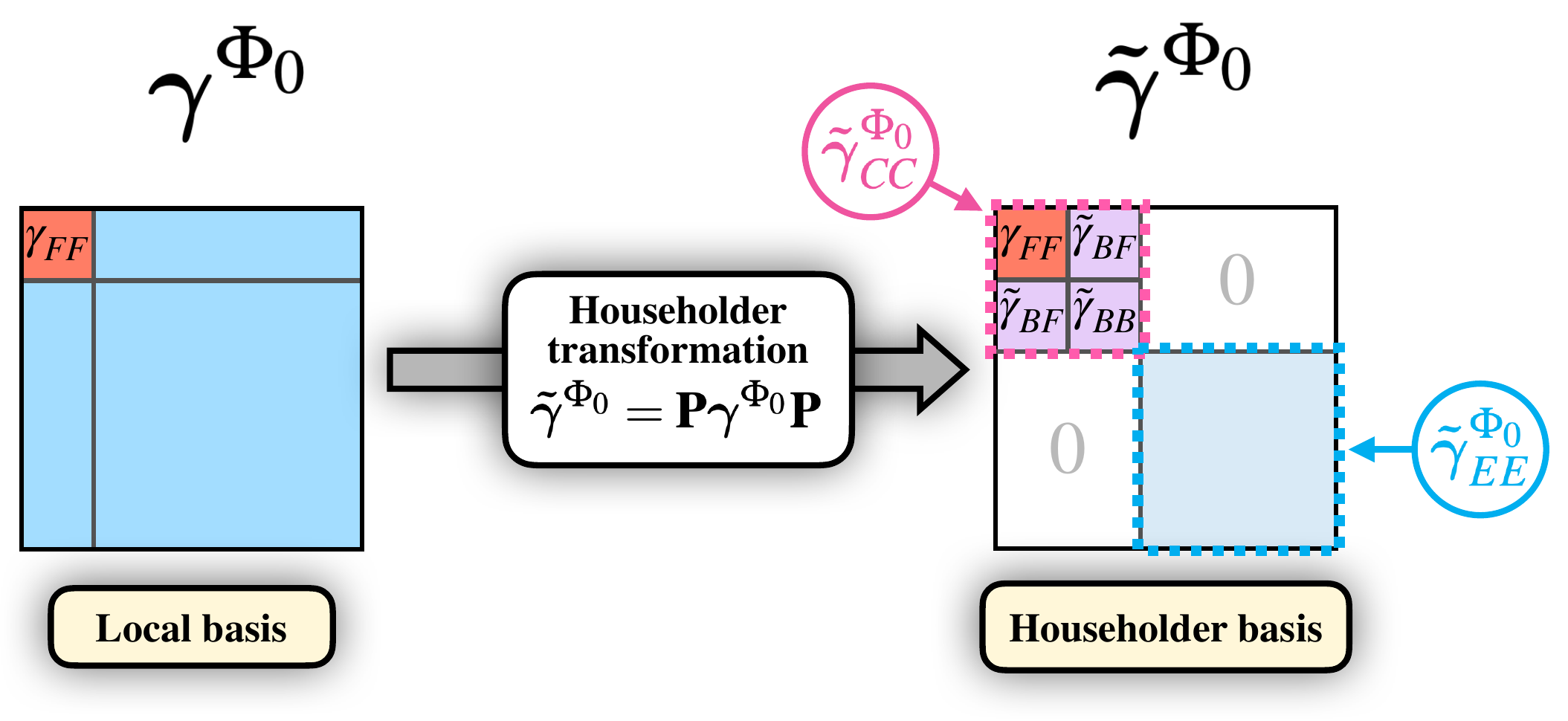}
\caption{Schematic representation of the
Householder transformation
on an idempotent 1-RDM of the mean-field state $\mykettight{\Phi_0}$. On the left: 1-RDM in the local basis, where $\gamma_{FF}$ denotes the fragment element. On the right: 1-RDM in the Householder basis, where the two decoupled sectors are delineated, the embedding cluster $\bgamt_{CC}^{\Phi_0}$ in the top left, and the cluster's environment $\bgamt_{EE}^{\Phi_0}$ in the bottom right.}   
\label{fig:Fig1}
\end{figure}

From a practical point of view, the Householder transformation, denoted $\bold{P}$, is a unitary transformation originally employed in the development of numerical algorithms such as QR factorization or tridiagonalization, to cite a few examples (see Refs.~\cite{householder_unitary_1958,wilkinson1962householder,martin1968householder}). In the context of ground state embedding, this approach has recently been highlighted by some of the present authors as a practical embedding technique~\cite{yalouz2022quantum,sekaran2021householder,sekaran2022local} that can straightforwardly yield optimal bath orbitals for a given local fragment. This tool operates by block-diagonalizing the reference 1-RDM given in Eq.~(\ref{eq:1-RDMdef}). To illustrate this, let us consider that the upper-left block of the 1-RDM, denoted $\gamma_{FF}$, contains a single-orbital fragment, as shown in the left part of Fig.~\ref{fig:Fig1}.
As explained in Refs~\cite{yalouz2021state,sekaran2021householder},
when transformed into the Householder basis like
\be\label{eq:bgam_Phi0_HHrep}
\bgamt^{\Phi_0} = \bP\bgam^{\Phi_0}\bP, 
\ee
 one can show\cite{yalouz2022quantum,sekaran2021householder} that the transformed 1-RDM $\bgamt^{\Phi_0}$ attains a block-diagonal form as illustrated in the right part of Fig.~\ref{fig:Fig1}. 
In this figure, the top left block of the 1-RDM is the Householder cluster block spanning the fragment and bath spin orbitals, and $\bgamt_{{E}{E}}^{\Phi_0}$ is the so-called Householder environment.
 This strict decoupling into two independent sub-blocks is due to the idempotency property of the mean-field 1-RDM, i.e. $\bgam^2=\bgam$ (see Ref.~\cite{yalouz2021state} for a mathematical proof).
Interestingly, another consequence of the idempotency is the integer trace
of the Householder cluster
\be
\Tr [ \bgamt_{{C}{C}}^{\Phi_0}]=1,
\ee 
which implies that the latter contains exactly $1$ electron per spin (equivalently $2$ electrons if we consider spatial orbitals). 
Consequently, these properties reveal that the Householder transformation gives rise to an exact fragment embedding at the mean-field level as here the original Slater determinant $\ket{\Phi_0}$ can be exactly factorized into two components,
\be
\mykettight{\Phi_0} =  \mykettight{\tilde{\Phi}_0^{{E}}} \mykettight{\tilde{\Phi}_0^{{C}}},
\ee
where $\mykettight{\tilde{\Phi}_0^{{E}}}$ and $\mykettight{\Phi_0^{{C}}}$ are wavefunctions of the Householder environment and Householder cluster (containing respectively $N-2$  and 2 electrons).
The resulting orbital space partitioning combined with the finite number of electrons allows a straightforward use of regular high-level wave function methods to describe the electronic correlation within the cluster
(e.g. active space configuration interaction).
This framework was employed in recent works as a starting point for the development of different flavours of DMET-like methods (see Refs.~\cite{yalouz2022quantum,sekaran2021householder,sekaran2022local}).

\subsection{Exact fragment embedding for ensembles of states at the non-interacting level}\label{sec:exact_embedding_NI_ensembles}

Going beyond the ground state, we will now demonstrate that the Householder transformation can also generate an exact fragment embedding adapted to ground and excited states simultaneously.
As ensembles are the original motivation of our approach, we will first provide more details on this new framework.
Then we will explain how to adapt the use of the Householder transformation to this new context to design an exact and democratic ensemble embedding at the non-interacting level. 

\subsubsection{ Original motivations: the ensemble framework }\label{sec:ensemble_embedding_original_motivations}

In electronic structure calculations, ensemble approaches are usually considered as an appropriate choice to study systems where several low-lying states need to be treated on an equal footing (e.g. in photochemical systems with degenerate states). 
Indeed, compared to state-specific approaches, ensemble-based methods have two main advantages: i) they provide an unbiased democratic treatment of multiple states, and ii) they are rigorously supported by the ensemble GOK variational principle~\cite{gross1988rayleigh}. 
To illustrate this, let us assume we are interested in the ground $\ket{\Psi_0}$ and first excited state $\ket{\Psi_1}$ (with respective energy $E_0$ and $E_1$) of generic many-electron Hamiltonian $\hat{H} = \hat{h} + \hat{g}$ where $\hat{h}$ and $\hat{g}$  are one- and two-electron operators.
Then, the GOK variational principle stipulates that any ensemble energy  $E^\bold{w} $ built from the weighted contribution of any couple of trial states, noted $\mykettight{\psi_0}$ and $\mykettight{\psi_1}$, will always follow
\be\label{eq:ens_Energy}
E^\bold{w} = w_0\brakettight{\psi_0}{\hat{H}}{\psi_0}
+w_1\brakettight{\psi_1}{\hat{H}}{\psi_1}
\geq
w_0 E_0 + w_1 E_1,
\ee 
where the two weights $\bw=(w_0,w_1)$ fulfill the constraints $w_0\geq w_1$ and $w_0 + w_1 = 1$.
Here, a strict equality is fulfilled if the trial states correspond to the exact
ground and excited states $\mykettight{\psi_0} = \mykettight{\Psi_0}$ and $\mykettight{\psi_1} = \mykettight{\Psi_1}$ of the targeted system.
Interestingly, the general definition of the ensemble energy $E^\bold{w} $ given in Eq.~(\ref{eq:ens_Energy}) holds true for any type of system, including noninteracting ones (or mean-field ones).
In the latter case, one can easily show that the (exact) ensemble energy is a functional of the ensemble 1-RDM noted  $\bgam^{\mathbf{w}}$  which is defined by
\be\label{eq:1-RDM_ensemble_MF}
\bgam^{\mathbf{w}} = w_0\bgam^{\Phi_0}
+ w_1\bgam^{\Phi_1},
\ee
where $\bgam^{\Phi_0}$ and $\bgam^{\Phi_0}$ are the 1-RDMs (as defined in Eq.~(\ref{eq:1-RDMdef})) associated to the exact ground and first excited state $\ket{\Phi_0}$ and $\ket{\Phi_1}$ of the non-interacting reference system.

The fact that the exact ensemble energy of a non-interacting system can be recovered only from its ensemble 1-RDM is a motivating starting point for the development of an exact fragment embedding for multiple states at the noninteracting level.
In this case, the ensemble 1-RDM naturally appears as the new ingredient to be manipulated.
However, additional care has to be paid here as ensemble 1-RDMs are generally not idempotent (i.e. $\bgam^{\mathbf{w}} \neq (\bgam^{\mathbf{w}})^2$).
Thus, it is not expected that a simple application of the  Householder transformation can lead to a block-diagonalization of the matrix (as explained in Ref.~\cite{sekaran2021householder}).
In the following, we will introduce an extension of this approach that can tackle this issue.



\subsubsection{Exact fragment embedding using series of Householder transformations }\label{sec:ensemble_embedding_two-state}


To fulfill the exact block-diagonalization of the non-idempotent ensemble 1-RDM, the strategy we propose consists in applying, not one, but a finite series of Householder transformations.
Such a technique is usually employed in tri-diagonalization procedures~\cite{householder1958generated,householder1958unitary} as a preprocessing step for computing eigenvalues of symmetric matrices~\cite{burden2011numerical} (see Appendix~\ref{appendix:clus_succ_householder} for more details). 
Concisely, it is performed by applying to a generic matrix $\bold{M}$ with a dimension $D$, a product
of $D-1$ Householder transformations $\mathbf{Q}  = \mathbf{P}^{(1)}\mathbf{P}^{(2)}\dots\mathbf{P}^{(D-1)}$ such that
\be
\label{eq:series_house}
\begin{aligned}
\tilde{\bold{M}}  = \mathbf{Q}^{ \dagger} \bold{M} \mathbf{Q}
\end{aligned}
\ee
is a tri-diagonal matrix. 
In the context of fragment embedding for ensembles, our objective is however a little bit different than tri-diagonalization: we want to reach a block-diagonal form for the ensemble 1-RDM $\bgam^\bold{w}$.
To reach this goal, a similar approach may be used in practice but with a truncated number of successive Householder transformations. 

\begin{figure}[t!]
    \centering \includegraphics[width=0.49\textwidth]{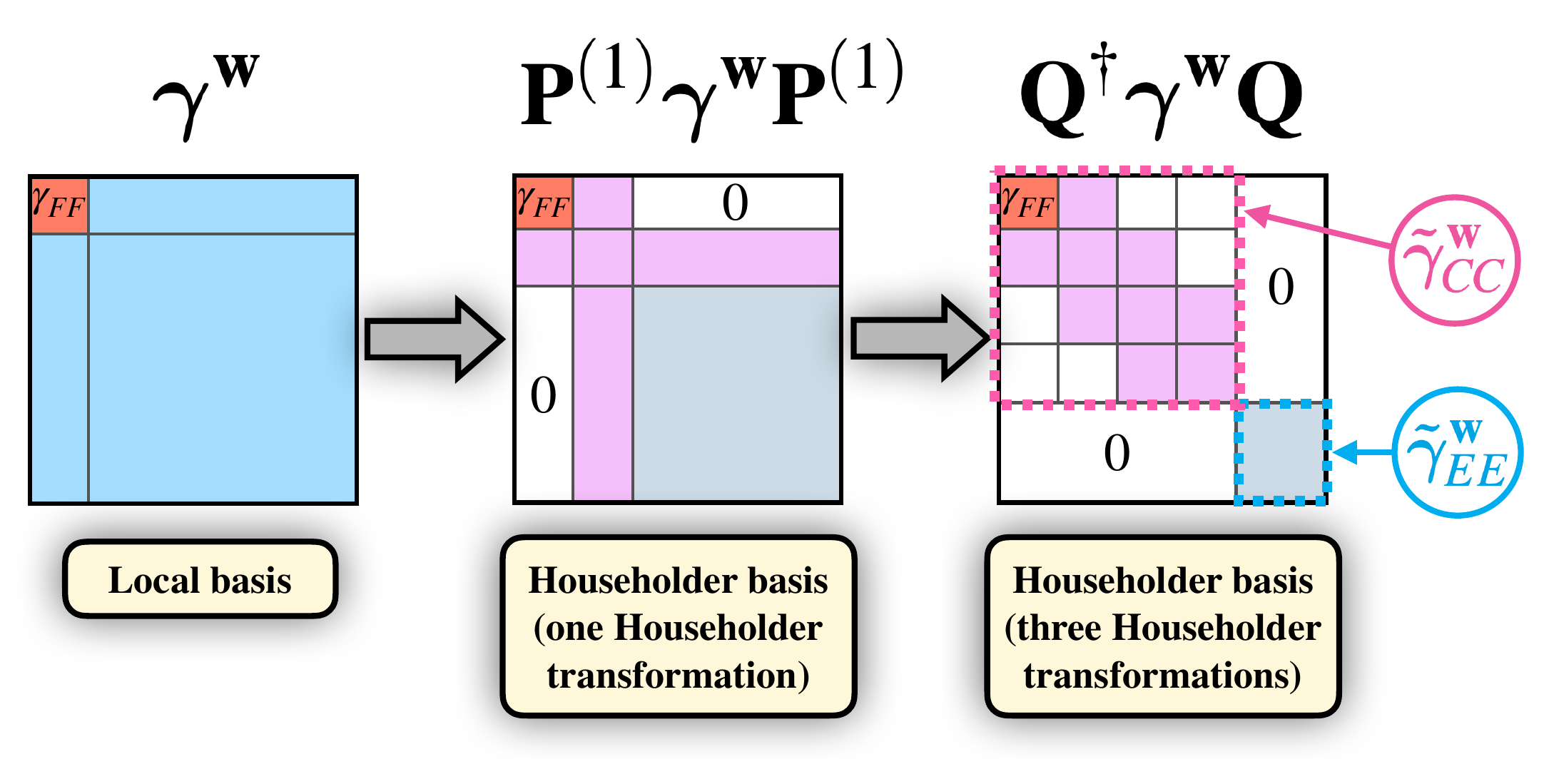}
    \caption{ Exact block-diagonalization of a two-state ensemble 1-RDM $\bgam^\bold{w}$ (as defined in Eq.~(\ref{eq:1-RDM_ensemble_MF})) based on three successive Householder transformations $\bQ = \bP^{(1)}\bP^{(2)}\bP^{(3)}$.}
    \label{fig:Fig2}
\end{figure}

To illustrate this, we will first consider the particular case of an ensemble 1-RDM built from only two states: a ground state $\ket{\Phi_0}$ and a first excited state $ \ket{\Phi_1}$ of a given non-interacting system (as given in Eq.~\eqref{eq:1-RDM_ensemble_MF}). 
In the eigen-orbital basis of the reference non-interacting Hamiltonian (or mean-field one), the ground state is a single Slater determinant with lowest orbitals doubly-occupied, while the first excited state is given by the HOMO-LUMO orbitals excitation such that
\be\label{eq:1-RDM_singly-excited}
\myket{\Phi_1} =
\dfrac{1}{\sqrt{2}}
\left(\sum_{\sigma=\uparrow,\downarrow}
\hat{a}_{ N/2+1,\sigma}^{\dagger}\hat{a}_{N/2,\sigma}\right)
\vert\Phi_0\rangle.
\ee
Starting from the associated ensemble 1-RDM $\bgam^{\bw}$ expressed in a local orbital basis, we can show that by applying three successive Householder transformations
\be\label{eq:ensemble_embedding_transformation_MF}
\bQ = \bP^{(1)}\bP^{(2)}\bP^{(3)},
\ee
the resulting transformed matrix
\be
\bgamt^{\bw} = \mathbf{Q}^{\dagger}\bgam^{\bw}\mathbf{Q},
\ee
is block diagonalized with a cluster block  fully decoupled form an environment block as exemplified in Fig.~\ref{fig:Fig2}. As shown in this figure, the dimension of the resulting cluster block is 
\be
\label{eq:dim_M2}
\dim (\bgamt_{{C}{C}}^{\bw}) = 4,
\ee
and we can also mathematically demonstrate (see Appendix~\ref{appendix:frac_rdm_general_math_proofs}) that the trace of this block is finite with
\be
\label{eq:trace_M2}
\Tr[\bgamt_{{C}{C}}^{\bw}] = 2.
\ee
This indicates that the cluster exactly contains $2$ electrons in total (equivalently $4$ electrons if we consider spatial orbitals). 
A closed cluster containing an integer number of electrons reveal that the resulting Householder orbital basis leads to an exact fragment embedding for the two-state ensemble.
Indeed, here both ground and excited state factorize into a cluster and environment parts such that
\be
\mykettight{\Phi_I} \overset{I=0,1}{=}
\mykettight{\Phi_0^{{E}}} \mykettight{\Phi_I^{{C}}},
\ee
where the cluster's environment $\mykettight{\Phi_0^{{E}}}$ is the
same fully-occupied Slater determinant for both states, since they have the same $N/2-1$ fully occupied orbitals.
This exact factorization of the two wavefunctions goes along the lines of what was obtained for a single state (see Section~\ref{sec:GS}).
Similarly, the orbital space partitioning allows a straightforward use of regular high-level methods to describe the electronic correlation within the cluster (e.g. in next steps of a DMET procedure).

The ability to construct an exact fragment embedding for a two-state ensemble marks a promising initial step, prompting inquiry into whether analogous characteristics persist for ensembles with $ M > 2 $ states, where the reference 1-RDM is
\be\label{eq:general_ens_1RDM}
\bold{\gamma}^\bold{w} = \sum_{I=0}^{ M-1 } w_I\  \bold{\gamma}^{\Phi_I}.
\ee
Numerical and analytical investigations actually revealed that an exact embedding can also be envisioned in this \textit{a priori} more involved case.
However, surprisingly, it turns out that the total number of states $M$ in the ensemble is not the predominant ingredient that will condition the embedding.
Indeed, it can be theoretically established that the properties of the final Householder cluster, such as dimensionality (i.e. $\dim(\bgamt_{{C}{C}}^{\bw})$) and the number of electrons it contains (i.e. $\Tr[\bgamt_{{C}{C}}^{\bw}]$), are actually only influenced by the fractionally occupied natural orbital subspace delineated within the reference ensemble 1-RDM $\bgam^\bold{w}$ (as defined in Eq.~(\ref{eq:general_ens_1RDM})).
Specifically, two pivotal factors of the ensemble 1-RDM will govern the final cluster's properties: the total count of fractionally occupied natural-orbitals, denoted as $N_{frac}^{orb}$, and the corresponding quantity of electrons, represented as $N_{frac}^{elec}$, inhabiting these natural orbitals. 
More precisely, we can show that the trace and the dimension of the cluster block $\bgamt_{{C}{C}}^{\bw}$ follow the linear behaviors: 
\be
\label{eq:lin_behavior}
\begin{cases}
 \dim(\bgamt_{{C}{C}}^{\bw}) &=   2 + N_{frac}^{orb}\\
\Tr[\bgamt_{{C}{C}}^{\bw}]    &=   1 + {N_{frac}^{elec}} / {2}.
 
\end{cases}
\ee   
For sake of conciseness, we will provide here a numerical proof of these properties, while directing interested readers to Appendix~\ref{appendix:frac_rdm_general_math_proofs} for a comprehensive mathematical proof of these behaviors.

To illustrate the laws given in Eq.~(\ref{eq:lin_behavior}), we have built an artificial ensemble 1-RDM $\gamma^\bold{w}$, wherein both parameters $N_{frac}^{orb}$ and \rev{${N_{frac}^{elec}}$} are systematically adjusted to delineate the characteristics of the fractionally occupied natural orbital subspace.
Following the specification of these parameters, a series of Householder transformations are applied iteratively on the reference matrix (as defined in Eq.~(\ref{eq:series_house})), ceasing when a decoupled cluster block is detected.
Subsequently, we evaluate the properties of the resulting cluster (for further technical details of these simulations, see Appendix~\ref{app:num_method}).
In Fig.~\ref{fig:Fig3}, we depict the number of Householder transformations necessary to generate an exact decoupling of the cluster along with the variation of the cluster block's dimension, $\dim(\bgamt_{{C}{C}}^{\bw})$, and its trace, $\Tr[\bgamt_{{C}{C}}^{\bw}]$, with respect to $N_{frac}^{orb}$ (and for different fixed values of \rev{$N_{frac}^{elec}=2,4,$} and $6$).
As depicted in the top panel of Fig.~\ref{fig:Fig3}, we see that the number of Householder transformations necessary to generate an independent cluster scales like $1 + N_{frac}^{orb}$, and the latter is totally independent from  ${N_{frac}^{elec}}$.
In the middle panel of Fig.~\ref{fig:Fig3},  similarly we note that the dimension of the cluster block, $\dim(\bgamt_{{C}{C}}^{\bw})$, only increases with ${N_{frac}^{orb}}$ and not \rev{${N_{frac}^{elec}}$}, in accordance with $\dim(\bgamt_{{C}{C}}^{\bw}) = 2 + N_{frac}^{orb}$ as defined in Eq.~(\ref{eq:lin_behavior}).
Conversely, in the bottom panel of Fig.~\ref{fig:Fig3}, we observe that the electron count within the cluster, represented by $\Tr[\bgamt_{{C}{C}}^{\bw}]$ is steady with the variation of ${N^{orb}_{frac}}$. However, reading vertically the plot, we see that the trace increases linearly with respect to ${N_{frac}^{elec}}$  following the behavior $\Tr[\bgamt_{{C}{C}}^{\bw}] = 1 + {N_{frac}^{elec}} / {2}$, as defined in Eq.~(\ref{eq:lin_behavior}). 
It's noteworthy that the series of data illustrated in Fig.~\ref{fig:Fig3} do not all initiate with the same ${N_{frac}^{orb}}$ value. 
This disparity stems from the constraint aimed at guaranteeing physically meaningful orbital occupation scenarios, which mandates that ${N_{frac}^{elec}} \rev{<} 2{N_{frac}^{orb}} $.  

The behaviors observed here evidence the possibility to exactly block-diagonalize a general ensemble 1-RDM with the use of a series of Householder transformations.
It is a proof that we can build an exact fragment embedding at the non-interacting level for $M$ states as the latter, in the Householder orbital basis, can be exactly factorized in a cluster and environment part as 
\be
\mykettight{\Phi_I} =
\mykettight{\Phi_0^{{E}}} \mykettight{\Phi_I^{{C}}}, \text{ for }  I= 0,1,\ldots, M-1.
\ee
However, this manipulation is not without consequences, as it inevitably leads to an increase in the dimensionality of the cluster, a parameter directly influenced by the quantity of fractionally occupied natural orbitals, $N_{frac}^{orb}$ (as detailed in Eq.~(\ref{eq:lin_behavior})). Naturally, this behavior may impose limitations on the strategy, especially when $N_{frac}^{orb}$ becomes comparable to the size $N$ of the system under study. Such a situation would arise as soon as we consider large ensembles composed of a large number of states decomposed onto a large number of orbitals. The investigation of such exotic situations is left for future work, as our current objective is more motivated by the ambition to target a few low-lying states requiring only a restricted number of orbitals at the non-interacting level.

To conclude this section, we emphasize
that so far, we have
focused on
exact embedding of
single-orbital fragments.
As discussed
in Section~\ref{eq:comparison_with_DMET},
by contrasting the application of SVD and Householder transformations in the generation of bath orbitals,
we can discern the underlying mechanisms driving the expansion of the
ensemble embedding cluster's dimension for
fragments of arbitrary size, thereby
generalizing the numerical results in Eq.~\eqref{eq:lin_behavior}.
Our interpretations provide compelling evidence that the enlargement of the cluster is a direct result of incorporating fractionally occupied natural orbitals from the ensemble 1-RDM.

\begin{figure}[t!]
\centering
\includegraphics[width=8.5cm]{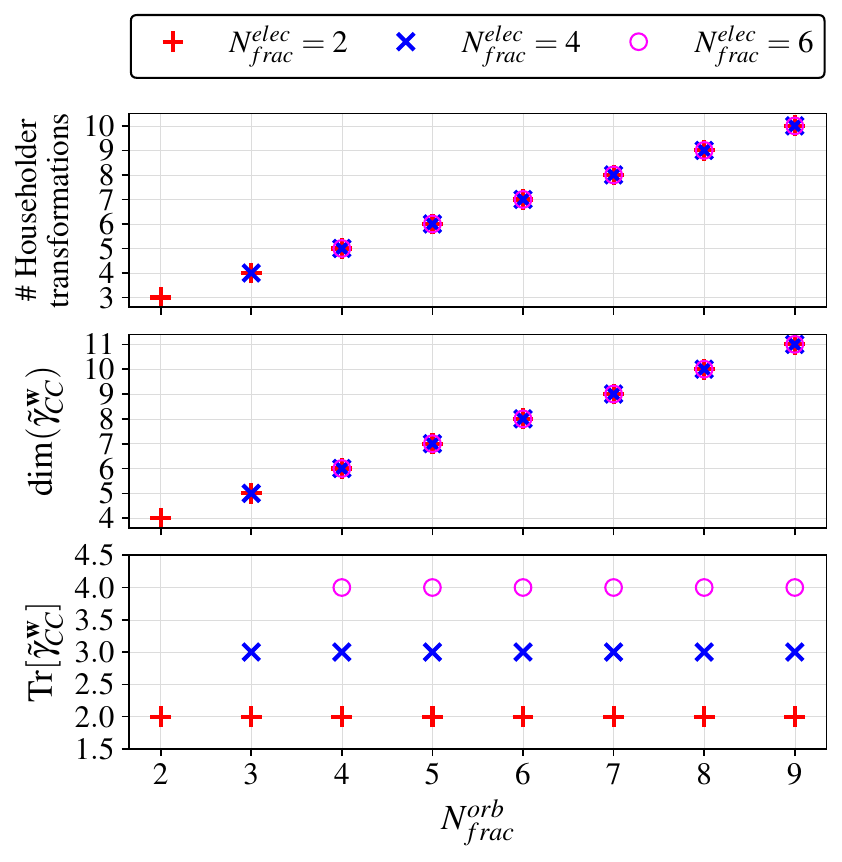}
\caption{
Evolution of the properties of the Householder cluster-block $\bgamt^\bold{w}_{CC}$ with the number of fractionally occupied orbitals $N^{orb}_{frac}$. \textbf{Top panel:} the number
of Householder transformations
required to generate a disentangled Householder cluster-block. \textbf{Middle panel:} dimension of the block.
\textbf{Bottom panel:} trace of the block, \textit{i.e.} number of electrons in the cluster.
To produce these results a reference 1-RDM $\bgam^\bold{w}$ is considered where we artificially modulate the $N_{frac}^{orb}$ and $N_{frac}^{elec}$ that defined the fractional occupied orbital space (see Appendix~\ref{app:num_method} for practical details).
}
\label{fig:Fig3}
\end{figure}

\section{Comparison with the DMET bath construction and extension to multi-orbital fragments}\label{eq:comparison_with_DMET}

In regular (ground-state) DMET~\cite{knizia2012density, wouters2016practical}, there are various (equivalent) ways to generate the (one-electron) quantum bath from an idempotent 1-RDM (simply referred to as density matrix in this section) that describes the entire system. One of them relies on the SVD of the fragment's environment-fragment block~\cite{zheng2018density,li2023multi}. It has been shown recently~\cite{sekaran2023unified} that this procedure generates the exact same one-electron bath subspace as the Householder transformation (or Block-Householder, if the to-be-embedded fragment contains several spin-orbitals) . In the following, we revisit the DMET bath construction for a non-interacting ensemble and show that, as expected from the previous section, the embedding can be made exact with a limited number (larger than for the ground-state problem) of bath spin-orbitals. The strategy applies also to ensemble mean-field calculations (the so-called ensemble density matrix Hartree--Fock approach in Ref.~\citenum{Cernatic2022}). By analogy with regular DMET, electron correlation can be introduced {\it a posteriori} within the embedding cluster, from which local ground- and excited-state properties can be computed, approximately (see Section~\ref{sec:Applications}). It should be possible to exactify formally the entire embedding procedure through the ensemble density-functional theory formalism~\cite{Cernatic2022}, by analogy with the ground-state local potential functional embedding theory~\cite{sekaran2022local}.
\rev{For the sake of clarity, the different one-electron representations that will be introduced step by step in the following, in order to achieve an ensemble embedding, are summarized in Table~\ref{table:one-electron_reps}.}
\\

\renewcommand{\arraystretch}{1.8}
\begin{table*}
\rev{
\begin{tabular}{ |c || c | c | c| } 
\hline
\bf Representation & \multicolumn{3}{|c|}{\bf Complete one-electron Hilbert space decompositions}\\ 
\hline
\hline
localized ($p,q,r,s,\ldots$) & fragment $F$  & fragment's environment $\overline{F}$  & $-$ \\
\hline
molecular ($\kappa$) & inactive spin-orbitals $\left\{\kappa_i\right\}\equiv {\rm In}$  & active spin-orbitals $\left\{\kappa_a\right\}\equiv {\rm Ac}$ & virtual spin-orbitals $\left\{\kappa_r\right\}$ \\
\hline
inactive embedding & fragment $F$ & inactive bath $B^{\rm In}$ & inactive cluster's environment $\mathcal{E}^{\rm In}$ \\
\hline
ensemble embedding & cluster $C=\underset{C^{\rm In}}{\underbrace{F\oplus B^{\rm In}}}\oplus\left(\hat{\mathds{1}}-\hat{P}_{C^{\rm In}}\right)\left\{\kappa_a\right\}$  & $C^{\perp}$ (not used) & $-$\\
\hline
\end{tabular}
\caption{Summary of the different one-electron representations involved in the construction of the complete ensemble embedding as discussed in Sec.~\ref{eq:comparison_with_DMET}.}
\label{table:one-electron_reps}
}
\end{table*}

\subsection{Ensemble energy within the molecular spin-orbital representation}

Let 
\be\label{eq:one_elec_hamil_localized_rep}
\hat{h}=\sum_{pq}h_{pq}\hat{c}_p^\dagger\hat{c}_q
\ee 
be the (non-interacting in the present case) second-quantized Hamiltonian written in a given orthonormal basis of localized orbitals (sites for lattice models). Once the Schr\"{o}dinger equation has been solved for a single electron, i.e.,  
\be
\hat{h}\myket{\kappa}=\varepsilon_\kappa\myket{\kappa},
\ee
where $\myket{\kappa}\equiv\hat{a}_\kappa^\dagger\myket{\rm vac}$ corresponds to a molecular spin-orbital with energy $\varepsilon_\kappa$ and $\myket{\rm vac}$ denotes the (zero-electron) vacuum state, we can, on the basis of the many-electron ensemble we want to study, split the one-electron Hilbert space 
into an {\it even} number $N^{\rm In}$ of inactive spin-orbitals $\kappa_i$ (the corresponding spatial orbitals are doubly-occupied in all the Slater determinants from which ground- and excited-state wavefunctions are constructed), $N^{\rm Ac}$ actives $\kappa_a$ (occupied in some of the Slater determinants from which ground and excited states are described), and the virtuals $\kappa_r$ (unoccupied in all the states that belong to the ensemble under study):
\be\label{eq:decomp_one_elec_Hilb_space_MO_rep}
\begin{aligned}
\left\{\myket{\kappa}\right\}
&
=\left\{\myket{\kappa_i}\right\}_{1\leq i\leq N^{\rm In} }\oplus\left\{\myket{\kappa_a}\right\}_{N^{\rm In}+1\leq a\leq N^{\rm In}+N^{\rm Ac} }
\\
&\quad\quad\oplus\left\{\myket{\kappa_r}\right\}_{r>N^{\rm In}+N^{\rm Ac}}.
\end{aligned}
\ee
The molecular spin-orbital energies are ordered as follows,
\be
\varepsilon_{\kappa_{k}}\leq \varepsilon_{\kappa_{k+1}}, 
\ee
and, by definition,
\be
\varepsilon_{\kappa_i} < \varepsilon_{\kappa_a} < \varepsilon_{\kappa_r}, \;\forall i,a,r. 
\ee
The many-electron eigenstates of $\hat{h}$ that form the ensemble have the following structure
\begin{subequations}
\begin{align}
\label{eq:wf_structure_2nd_quant}
\myket{\Phi_I}&=\left(\hat{\Phi}^{\rm In}\right)^\dagger\left(\hat{\Phi}^{\rm Ac}_I\right)^\dagger\myket{\rm vac}=\left(\hat{\Phi}^{\rm Ac}_I\right)^\dagger\left(\hat{\Phi}^{\rm In}\right)^\dagger\myket{\rm vac}
\\
\label{eq:PhiIn_PhiAc_tensor_prod}
&\equiv \myket{{\Phi}^{\rm In}{\Phi}^{\rm Ac}_I},
\end{align}
\end{subequations}
where  
\be
\left(\hat{\Phi}^{\rm In}\right)^\dagger=\prod^{N^{\rm In}}_{i=1}\hat{a}_{\kappa_i}^\dagger
\ee
creates the closed-shell Slater determinant
\be
\myket{\Phi^{\rm In}}=\left(\hat{\Phi}^{\rm In}\right)^\dagger\myket{\rm vac}
\ee
that describes the $N^{\rm In}$ inactive electrons to which the regular DMET bath construction will be applied later on. 
The $\mathcal{N}^{\rm Ac}_e$-electron states $\myket{\Phi^{\rm Ac}_I}:=\left(\hat{\Phi}^{\rm Ac}_I\right)^\dagger\myket{\rm vac}$, which are a priori {\it not} single Slater determinants, unlike $\Phi^{\rm In}$, are constructed from the active spin orbitals whose occupation in $\myket{\Phi^{\rm Ac}_I}$ depends on the state $I$ under consideration within the ensemble:
\be
\myket{\Phi^{\rm Ac}_I}\in\left\{\prod^{N^{\rm In}+N^{\rm Ac}}_{a=N^{\rm In}+1}\left(\hat{a}_{\kappa_a}^\dagger\right)^{n_{\kappa_a}}\myket{\rm vac}\right\},
\ee
where $n_{\kappa_a}\in\{0,1\}$ and $\sum^{N^{\rm In}+N^{\rm Ac}}_{a=N^{\rm In}+1}n_{\kappa_a}=\mathcal{N}^{\rm Ac}_e$.
From the Hamiltonian representation in the molecular spin-orbital basis,
\begin{subequations}
\begin{align}
\hat{h}&=\sum^{N^{\rm In}}_{i=1}\varepsilon_{\kappa_i} \hat{a}_{\kappa_i}^\dagger \hat{a}_{\kappa_i}+\sum^{N^{\rm In}+N^{\rm Ac}}_{a=N^{\rm In}+1}\varepsilon_{\kappa_a} \hat{a}_{\kappa_a}^\dagger \hat{a}_{\kappa_a}+\sum_{r>N^{\rm Ac}+N^{\rm In}}\varepsilon_{\kappa_r} \hat{a}_{\kappa_r}^\dagger \hat{a}_{\kappa_r}
\\
&=:\hat{h}^{\rm In}+\hat{h}^{\rm Ac}+\sum_{r>N^{\rm Ac}+N^{\rm In}}\varepsilon_{\kappa_r} \hat{a}_{\kappa_r}^\dagger \hat{a}_{\kappa_r},
\end{align}
\end{subequations}
we can finally evaluate the (minimizing GOK~\cite{gross1988rayleigh}) ensemble energy as follows, 
\begin{subequations}
\begin{align}
E^{\bw}&
=\sum_{I\geq 0}w_I\mybraket{\Phi_I}{\hat{h}}{\Phi_I}
\\
&=\sum_{I\geq 0}w_I\left(\mybraket{\Phi_I}{\hat{h}^{\rm In}}{\Phi_I}+\mybraket{\Phi_I}{\hat{h}^{\rm Ac}}{\Phi_I}\right),
\end{align}
\end{subequations}
thus leading to (see Eq.~(\ref{eq:wf_structure_2nd_quant}))
\be
E^{\bw}=\sum_{I\geq 0}w_I\left(\mybraket{\Phi^{\rm In}}{\hat{h}^{\rm In}}{\Phi^{\rm In}}+\mybraket{\Phi^{\rm Ac}_I}{\hat{h}^{\rm Ac}}{\Phi^{\rm Ac}_I}\right),
\ee
or, equivalently,
\be\label{eq:final_exp_MO_rep_ens_ener}
E^{\bw}=\sum_{I\geq 0}w_I\left(\mybraket{\Phi^{\rm In}}{\hat{h}}{\Phi^{\rm In}}+\mybraket{\Phi^{\rm Ac}_I}{\hat{h}}{\Phi^{\rm Ac}_I}\right).
\ee
From the above energy expression, we first realize that the regular DMET algorithm (or, equivalently, the Householder transformation~\cite{sekaran2023unified}) can be used straightforwardly to evaluate the inactive part of the ensemble energy (first term on the right-hand side of Eq.~(\ref{eq:final_exp_MO_rep_ens_ener})), simply because it is described by a single Slater determinant. Increasing the embedding cluster's size and the number of electrons it contains, in order to describe the active part of the ensemble energy (second term on the right-hand side of Eq.~(\ref{eq:final_exp_MO_rep_ens_ener})), will provide a rationale for the conclusions drawn in Section~\ref{sec:ensemble_embedding_two-state} (see Eq.~(\ref{eq:lin_behavior})). This is discussed in the next section. 


\subsection{Embedding of the inactive energy and enlargement of the embedding cluster with the actives}\label{sec:embedding_inactives_and_actives}

\rev{\subsubsection{Conventional DMET bath construction applied to the inactive electrons}}

We now consider an alternative decomposition of the one-electron Hilbert space into the to-be-embedded fragment spin-orbital subspace $F$ of dimension $L_F$ and its complement $\overline{F}$, which corresponds to the complete and true environment of the fragment in the system under study:
\be
\left\{\myket{\kappa_i}\right\}\oplus\left\{\myket{\kappa_a}\right\}\oplus\left\{\myket{\kappa_r}\right\}=F\oplus\overline{F}.
\ee 
Unlike in Section~\ref{sec:Theor}, we consider the more general case where multiple (spatial) orbitals can be embedded (i.e., $L_F>2$).
On that basis, a third (so-called embedding) representation can be constructed by applying the regular DMET bath construction to the single-determinantal inactive wavefunction $\Phi^{\rm In}$, thus leading to the following decomposition (see, for example, Ref.~\citenum{sekaran2023unified} for further details),  
\be\label{eq:decomp_embedding}
F\oplus\overline{F}=\underset{C^{\rm In}}{\underbrace{F\oplus B^{\rm In}}}\oplus \mathcal{E}^{\rm In},
\ee
where the one-electron bath subspace $B^{\rm In}$, whose dimension equals that of the fragment ($\dim B^{\rm In}= \dim F = L_F$), will be referred to as {\it inactive bath} because it is constructed from $\Phi^{\rm In}$ or, more precisely, from the (one-electron reduced) density matrix ${\bgam}^{\rm In}$ of $\Phi^{\rm In}$ written in the original localized representation (see Eq.~(\ref{eq:one_elec_hamil_localized_rep})):  
\be
{\gamma}^{\rm In}_{pq}=
\mybraket{\Phi^{\rm In}}{\hat{c}_p^\dagger\hat{c}_q}{\Phi^{\rm In}}
.
\ee
Indeed, a simple basis of $B^{\rm In}$ (which can be orthonormalized via the SVD of the $\overline{F}F$ block \rev{${\bgam}^{\rm In}_{\overline{F}F}$} of ${\bgam}^{\rm In}$, for example) is generated as follows~\cite{sekaran2023unified},
\be\label{eq:def_inactive_bath}
B^{\rm In}=\left\{\sum_{q\in \overline{F}}\gamma^{\rm In}_{pq}
\hat{c}^\dagger_q\myket{{\rm vac}}
\right\}_{p\in F}.
\ee
\rev{Note that, as discussed in detail in Ref.~\citenum{sekaran2023unified}, solving the full SVD problem for the matrix $A={\bgam}^{\rm In}_{\overline{F}F}$ consists in diagonalizing the $L_{\overline{F}}\times L_{\overline{F}}$ (we denote $L_{\overline{F}}=\dim \overline{F} = 2L - L_F> L_F$) Hermitian matrix $AA^\dagger$, which provides, ultimately, a complete orthonormal basis for the true fragment's environment $\overline{F}$. The first (trivial) set of solutions, for which the eigenvalues equal zero, is the subspace (denoted $\mathcal{E}^{\rm In}$ in Eq.~(\ref{eq:decomp_embedding})) that is orthogonal to all the column vectors of $A$. If needed, the description of this subspace requires the computation of an orthonormal basis of dimension $L_{\overline{F}}-L_F$. The second set of solutions, which corresponds to the (inactive in this context) quantum bath $B^{\rm In}$, is obtained by diagonalizing $AA^\dagger$ within the subspace of the column vectors of $A$, as readily seen from Eq.~(\ref{eq:def_inactive_bath}). This is equivalent to diagonalizing the $L_F\times L_F$ Hermitian matrix $A^\dagger A$ and the resulting eigenvalues are the squared singular values of $A$. Localized and embedding representations can then be connected through a unitary transformation (that applies to the {\it entire} one-electron Hilbert space and is determined from the two orthonormal sets of SVD solutions and the fragment~\cite{sekaran2023unified}): 
} 
\be\label{eq:inactive_elec_embedding_unit_trans}
\bgam^{\rm In} \rightarrow \tilde{\bgam}^{\rm In}=\left(\mathcal{U}^{\rm In}\right)^\dagger \bgam^{\rm In} \mathcal{U}^{\rm In} 
.
\ee
\rev{Alternatively, a Householder transformation~\cite{AML99_Rotella_Block_Householder_transf,sekaran2023unified} can be applied to $\bgam^{\rm In}$. While providing the same subspaces $B^{\rm In}$ and $\mathcal{E}^{\rm In}$ as the SVD~\cite{sekaran2023unified}, it generates automatically (and analytically) a complete orthonormal basis for {\it both} subspaces just from the diagonalization of an auxiliary (determined from $A$) $L_F\times L_F$ Hermitian matrix. This is not automatic in the SVD, as seen from the above discussion. In practical (approximate) DMET calculations~\cite{wouters2016practical}, however, $\mathcal{E}^{\rm In}$ is not needed because the electronic structure problem is projected onto the embedding cluster subspace $C^{\rm In}$ (see Eq.~(\ref{eq:decomp_embedding})).}\\

\rev{Returning to the unitary transformation of Eq.~(\ref{eq:inactive_elec_embedding_unit_trans}), it leaves}, by construction, the fragment unchanged and, as a consequence of Eq.~(\ref{eq:def_inactive_bath}) {\it and} the idempotency of ${\bgam}^{\rm In}$~\cite{sekaran2023unified}, it ensures the strict disentanglement of the inactive embedding cluster $C^{\rm In}$ (fragment+inactive bath) from its environment $\mathcal{E}^{\rm In}$ (see Eq.~(\ref{eq:decomp_embedding})), i.e.,
\be\label{eq:decoupling_1RDM}
\tilde{\gamma}^{\rm In}_{pq}\overset{p\in\mathcal{E}^{\rm In}}{\underset{q\in C^{\rm In}}{=}}0. 
\ee
It also guarantees that the \rev{{\it inactive}} cluster contains as many electrons as the number of spin-orbitals in the fragment~\cite{sekaran2023unified}:
\be\label{eq:nbr_electrons_inactive_cluster}
\sum_{p\in C^{\rm In}}\tilde{\gamma}^{\rm In}_{pp}=L_F.
\ee
\rev{Therefore, like in regular (pure) ground-state DMET, the filling of the inactive embedding cluster equals $L_F/(2L_F)=1/2$. Of course, at this point, we have not embedded the active electrons yet. The complete ensemble embedding will be presented in Sec.~\ref{sec:embedding_active_electrons}.}
\rev{Nevertheless, we can already evaluate} the inactive part of the ensemble energy (first term on the right-hand side of Eq.~(\ref{eq:final_exp_MO_rep_ens_ener})) in the embedding representation as follows (we use real algebra for simplicity),
\begin{subequations}
\begin{align}
\mybraket{\Phi^{\rm In}}{\hat{h}}{\Phi^{\rm In}}
&=\Tr\left[h\bgam^{\rm In}\right]=\Tr\left[\tilde{h}\bgamt^{\rm In}\right]
\\
\label{eq:sep_cluster_clusterenv_inactive_energy}
&=\sum_{pq\in C^{\rm In}}\tilde{h}_{pq}\tilde{\gamma}^{\rm In}_{pq}+\sum_{pq\in \mathcal{E}^{\rm In}}\tilde{h}_{pq}\tilde{\gamma}^{\rm In}_{pq},
\end{align}
\end{subequations}
where $h\equiv \left\{h_{pq}\right\}$ and $\tilde{h}=\left(\mathcal{U}^{\rm In}\right)^\dagger h\mathcal{U}^{\rm In}$. As the idempotency of $\bgam^{\rm In}$ implies that of its $C^{\rm In}C^{\rm In}$ block (because of Eq.~(\ref{eq:decoupling_1RDM})), it comes from Eq.~(\ref{eq:nbr_electrons_inactive_cluster}) that the inactive cluster part of the energy (first term on the right-hand side of Eq.~(\ref{eq:sep_cluster_clusterenv_inactive_energy})) can be evaluated from a single-determinantal $L_F$-electron wavefunction $\Phi^{C^{\rm In}}$. Note that, in practice, $\Phi^{C^{\rm In}}$ would be determined by diagonalizing the projection of $\hat{h}$ onto the $L_F$-electron Fock subspace that is generated from the inactive embedding cluster. In this representation, $\Phi^{C^{\rm In}}$ is written as a multi-determinantal wavefunction. In practical DMET computations, the electronic repulsion within the fragment (possibly the bath too) would also be taken into account, thus leading, in this case, to an (approximate) embedding scheme where $\Phi^{C^{\rm In}}$ would become a correlated wavefunction. Returning to the non-interacting case, in which the embedding is exact, we finally obtain the following expression for the inactive energy,
\be\label{eq:inac_ener_split_cluster_and_env}
\mybraket{\Phi^{\rm In}}{\hat{h}}{\Phi^{\rm In}}=\mybraket{\Phi^{C^{\rm In}}}{\hat{h}}{\Phi^{C^{\rm In}}}+\sum_{pq\in \mathcal{E}^{\rm In}}\tilde{h}_{pq}\tilde{\gamma}^{\rm In}_{pq}.
\ee
\rev{Note that, in the diagonalizing one-electron representation of the inactive embedding cluster contribution (first term on the right-hand side of Eq.~(\ref{eq:inac_ener_split_cluster_and_env})),} the inactive embedding cluster subspace will be split into occupied-in-$\Phi^{C^{\rm In}}$ spin-orbitals (which belong to the inactive subspace) and unoccupied-in-$\Phi^{C^{\rm In}}$ spin-orbitals, which belong to the active+virtual subspace. \rev{As a final comment and reminder, so far, we have simply applied the regular DMET approach to the inactive electrons.}

\rev{\subsubsection{From the pure single-determinantal inactive to the complete ensemble state embedding}\label{sec:embedding_active_electrons}}

In order to recover the active contributions to the ensemble energy (second term on the right-hand side of Eq.~(\ref{eq:final_exp_MO_rep_ens_ener})), we need to complement the inactive embedding cluster with the active spin-orbital subspace (from which the projection onto the inactive cluster is removed, via the projection operator $\hat{P}_{C^{\rm In}}$).
Thus, we obtain an enlarged and definitive embedding cluster subspace 
\be\label{eq:definitive_emb_cluster_decomp}
{C}=C^{\rm In}\oplus \left\{\left(\hat{\mathds{1}}-\hat{P}_{C^{\rm In}}\right)\myket{\kappa_a}\right\}_{N^{\rm In}+1\leq a\leq N^{\rm In}+N^{\rm Ac}}
\ee
from which the exact ensemble energy can be evaluated as follows (note that the weights sum up to 1),
\be\label{eq:nonint_ens_ener_cluster_plus_env}
E^{\bw}=\sum_{I\geq 0}w_{I}\mybraket{\Phi_I^{C}}{\hat{h}}{\Phi_I^{C}}
+
\sum_{pq\in \mathcal{E}^{\rm In}}\tilde{h}_{pq}\tilde{\gamma}^{\rm In}_{pq},
\ee
where, with the notation of Eq.~(\ref{eq:PhiIn_PhiAc_tensor_prod}), the embedded ground and excited states read 
\be\label{eq:embedded_states_tensor_prod}
\myket{\Phi_I^{{C}}}\equiv \myket{\Phi^{{C}^{\rm In}}\Phi_I^{\rm Ac}}.
\ee
As readily seen from Eqs.~(\ref{eq:decomp_embedding}) and (\ref{eq:nonint_ens_ener_cluster_plus_env}), the cluster ensemble energy (first sum on the right-hand side of Eq.~(\ref{eq:nonint_ens_ener_cluster_plus_env})) is the only energy contribution we need to evaluate (variationally) to retrieve ground- and excited-state properties (i.e., reduced density matrices) within the fragment subspace, which is orthogonal to the inactive embedding cluster's environment subspace $\mathcal{E}^{\rm In}$. This is simply achieved by diagonalizing the projection of the Hamiltonian $\hat{h}$ onto the space of Slater determinants that is generated by distributing $L_F+\mathcal{N}^{\rm Ac}_e$ electrons (the first $L_F$ electrons come from the inactive embedding cluster, as readily seen from Eq.~(\ref{eq:embedded_states_tensor_prod})) among $2L_F+N^{\rm Ac}$ spin-orbitals (the first $2L_F$ spin-orbitals come from the fragment and the inactive bath, as readily seen from Eqs.~(\ref{eq:decomp_embedding}) and (\ref{eq:definitive_emb_cluster_decomp})). As we did not specify the number $L_F$ of fragment spin-orbitals that are embedded, we generalize the key result of Eq.~(\ref{eq:lin_behavior}), which is recovered in the particular case where a single (spatial) orbital is embedded, i.e., $L_F=2$. In this case, if we use the notations of Eq.~(\ref{eq:lin_behavior}), the dimension of the complete embedding cluster $C$ (including both spin states) is $4+N^{\rm Ac}=2(2+N_{frac}^{orb})$, and the total number of electrons in the cluster is $2+\mathcal{N}^{\rm Ac}_e=2(1+(N_{frac}^{elec}/2))$, \rev{which is in perfect agreement with Fig.~\ref{fig:Fig3}.}\\ 

\rev{Note finally that the filling of the complete ensemble embedding cluster equals
\be
\dfrac{L_F+\mathcal{N}^{\rm Ac}_e}{2L_F+N^{\rm Ac}}=\dfrac{1}{2}\dfrac{\left(L_F+N_{frac}^{elec}\right)}{L_F+N_{frac}^{orb}}
=\dfrac{1}{2}-\dfrac{1}{2}\dfrac{\left(N_{frac}^{orb}-N_{frac}^{elec}\right)}{L_F+N_{frac}^{orb}}.
\ee
Therefore, unlike in regular ground-state DMET, where $N_{frac}^{orb}=N_{frac}^{elec}=0$, the ensemble embedding cluster is less than half-filled if the number of active orbitals is strictly larger than the number of active electrons ($N_{frac}^{orb}>N_{frac}^{elec}$). It will be more than half-filled if, on the contrary, $N_{frac}^{orb}<N_{frac}^{elec}$. Half filling is actually recovered in the special case where the active orbital space is itself half-filled ($N_{frac}^{orb}=N_{frac}^{elec}$).
}


\subsection{Summary and key conclusions of the comparison}

The characteristics of the (single-orbital fragment) embedding cluster (i.e., its dimension and the number of electrons it contains), as constructed mathematically in Section~\ref{sec:ensemble_embedding_two-state} from successive ensemble density matrix functional Householder transformations, have been recovered from a different perspective, by extending the regular DMET bath construction to (GOK) ensembles. For that purpose, we first embedded a fragment (of arbitrary dimension $L_F$) in the presence of the inactive electrons, thus leading to an inactive embedding cluster of dimension $2L_F$ that contains $L_F$ electrons. The latter cluster was then complemented by $N^{\rm Ac}$ active spin-orbitals in which $\mathcal{N}^{\rm Ac}_e$ electrons are originally (i.e., at the full-size level of calculation) distributed to create the ensemble under study. We finally obtain a cluster of dimension $2L_F+N^{\rm Ac}$ (total number of spin-orbitals) that contains $L_F+\mathcal{N}^{\rm Ac}_e$ electrons. The procedure is exact for non-interacting (or mean-field) ensembles and applies to both single- and multi-orbital fragment embeddings. As discussed in further details in Section~\ref{sec:Applications}, its practical extension to the interacting problem (for which the embedding becomes actually useful) consists in introducing the electronic repulsion within the cluster, by analogy with regular ground-state DMET. The embedding becomes approximate in this case.

Let us finally stress that, even though the present \rev{(SVD-based)} DMET-like construction of the ensemble embedding cluster is simpler in its mathematical formulation than the composition of Householder transformations (as introduced in Section~\ref{sec:exact_embedding_NI_ensembles}), the latter has the main advantage \rev{over the SVD} (that is not exploited in the present work) of providing automatically a complete orthonormal embedding basis for the full one-electron Hilbert space (i.e., for the embedding cluster {\it and} its environment). \rev{This statement actually holds for both pure single-determinantal (a single Householder transformation is sufficient in this case) and ensemble states. A detailed explanation is given after Eq.~(\ref{eq:def_inactive_bath}) for the pure state that describes the inactive electrons but the same argument can be used for ensembles, as seen from Secs.~\ref{sec:ensemble_embedding_two-state} and \ref{sec:embedding_inactives_and_actives}.} Future developments where, for example, multi-reference perturbation theory could be applied on top of the embedding calculation~\cite{sekaran2021householder} (where the cluster is treated as a closed system), in order to recover correlations that are external to the cluster, would obviously benefit from this feature.

\section{Applications: A SINGLE-SHOT EMBEDDING STRATEGY FOR TWO-STATE ENSEMBLES}\label{sec:Applications}

\subsection{General description of the method}\label{sec:Applications_general_description}


In the following, we present a first
step toward
practical implementations
of fragment embedding for multiple states.
While the findings of numerical
study in Section~\ref{sec:ensemble_embedding_two-state} and the above comparison
with DMET
allude to the possibility of developing
a general scheme for targeting an arbitrary
number of states, we propose in this section
a simple strategy for calculating
the energies of
ground and first excited singlet states, 
inspired by the single-shot embedding
from DMET-like methods~\cite{wouters2016practical,sekaran2021householder,yalouz2022quantum}.

In essence, the strategy consists of utilizing
the bath orbitals of the noninteracting
Householder
cluster as the building
blocks of the
many-electron basis for
correlated embedding cluster wavefunctions.
For each fragment
in the local basis, the
bath orbitals are extracted
with the previously described
technique of applying successive Householder transformations
on the ensemble 1-RDM until
block-diagonalization. 
Afterwards, the embedding Hamiltonian
$\hat{H}^{C}$
is constructed by projecting 
the many-electron Hamiltonian $\hat{H}$ into
the Hilbert space of the ``fragment+bath"
cluster block. In DMET, different
versions of $\hat{H}^{C}$ are
possible depending on the extent of
electron interactions.
In the so-called
Interacting Bath (IB) formalism,
electron interactions appear in both the fragment
and bath orbitals, while in the
Noninteracting Bath (NIB)
formalism, electrons interact only on the fragment~\cite{wouters2016practical,wouters2017five,sekaran2021householder,marecat2023versatile}. In our strategy, we use the IB formalism.
In addition,
electron interactions between
the Householder cluster and the occupied orbitals inside the Householder
environment are incorporated in a mean-field way into
the one-electron part of $\hat{H}^{C}$.
At this point,
one may proceed to compute 
the correlated cluster 
wavefunctions $\mykettight{\Psi_I^{C}}$
by diagonalizing
$\hat{H}^{C}$ with
a high-level wavefunction method, in our
case the Full Configuration Interaction (FCI) method.
However, it is usually
observed that local orbital occupations
on fragments
change drastically when introducing
electron interactions inside the cluster.
As a consequence, the total number of electrons
when summed up from different fragments may
not match the true number of electrons in the whole system.
A simple solution in DMET is to optimize the
global number of electrons by
adding a chemical potential $\mu$
on the fragment, thereby enabling
the fluctuation of electrons
between the fragment and bath in
each embedding cluster~\cite{wouters2016practical,wouters2017five}.
Here, we adopt
a similar approach, with
a slightly more flexible optimization scheme.
For each fragment, with orbital indexed "$p$",
we solve the following
Schr\"{o}dinger equation,
\be
\left(\hat{H}^{C_p}-\mu_p \hat{n}_p\right)\mykettight{\Psi_I^{C_p}} = E_I^{C_p}\mykettight{\Psi_I^{C_p}},
\ee
to obtain the correlated
cluster wavefunctions $\mykettight{\Psi_I^{C_p}}$.
In the above equation, $\mu_p$ is the chemical potential and ${\hat{n}_{p}=\sum_{\sigma=\uparrow,\downarrow}\cdop_{p,\sigma}\cop_{p,\sigma}}$ the
density operator on fragment $p$, and $E_I^{C_p}$ is the
cluster energy (not used in the final energy evaluations).
The chemical potentials $\{\mu_p\}_{p=1}^L$,
which are in our calculations allowed
to differ across fragments, are adjusted to
correct for the right
total number of electrons
in both the ground and the
first excited state, by minimizing the following
cost function,
\be\label{eq:Ensemble_embedding_CF}
{\rm CF}(\{\mu_p\}) = \sum_{I=0,1}\left(
\sum_{p=1}^{L}
\brakettight{\Psi_I^{{C}_p}(\mu_p)}{\hat{n}_p}{\Psi_I^{{C}_p}(\mu_p)} - N
\right)^2.
\ee
After the optimization
of chemical potentials, the 1- and the
2-RDMs of the ground and excited state
wavefunctions of local clusters
are assembled to estimate the
total energies and other properties of the complete system. For the calculation
of energies, we make use of
the so-called democratic partitioning
approach, which has been extensively
discussed in previous works on DMET and related 
orbital-based quantum embedding methods~\cite{wouters2016practical,wouters2017five,nusspickel2023effective}.
In brief, the electronic energy
for each individual state is calculated as
\be\label{eq:individual_elec_energies}
E_I = \sum_{pq}h_{pq}D_{pq}^{\Psi_I} + \sum_{pqrs}g_{pqrs}\Gamma_{pqrs}^{\Psi_I},
\ee
where $h_{pq}$ and $g_{pqrs}$ are the
one- and two-electron integrals in the local basis, and
$D_{pq}^{\Psi_I}$ and $\Gamma_{pqrs}^{\Psi_I}$ are the
spin-free 1- and 2-RDMs, respectively. The latter
are estimated from embedding calculations on
different fragments
in a ``democratic"
fashion as follows,
\be
D_{pq}^{\Psi_I} = \dfrac{1}{2}\sum_{x=p,q}\,\sum_{\sigma=\uparrow,\downarrow}
\langle\Psi_I^{C_x}|\cdop_{p,\sigma}\cop_{q,\sigma}|\Psi_I^{C_x}\rangle,
\ee
\be
\Gamma_{pqrs}^{\Psi_I} = \dfrac{1}{4}\sum_{x=p,q,r,s}\,\sum_{\sigma,\sigma'=\uparrow,\downarrow}
\langle\Psi_I^{C_x}|\cdop_{p,\sigma}\cdop_{q,\sigma'}\cop_{s,\sigma'}\cop_{r,\sigma}|\Psi_I^{C_x}\rangle.
\ee
In \textit{ab initio} systems, the nuclear repulsion
energy (a state-independent term) is added to the electronic energies in Eq.~\eqref{eq:individual_elec_energies}.

We apply this two-state ensemble embedding strategy
in the computation of
energy curves in two 
simple
illustrative examples, the first one
being a finite Hubbard lattice model, and the second one an
\textit{ab initio} system of Hydrogen atoms. For benchmarking
the results of embedding calculations,
the FCI approach was used to extract
energy levels of ground and first-excited states. In our calculations, the
construction of fermionic 
creation and annihilation operators, and the subsequent construction and diagonalization of FCI and embedding
Hamiltonians were carried out using the python package QuantNBody~\cite{yalouz2022quantnbody}.

\subsection{Hubbard ring with site potential asymmetry and \rev{bond} length alternation}

\begin{figure}[!t]
\includegraphics[width=0.53\textwidth]{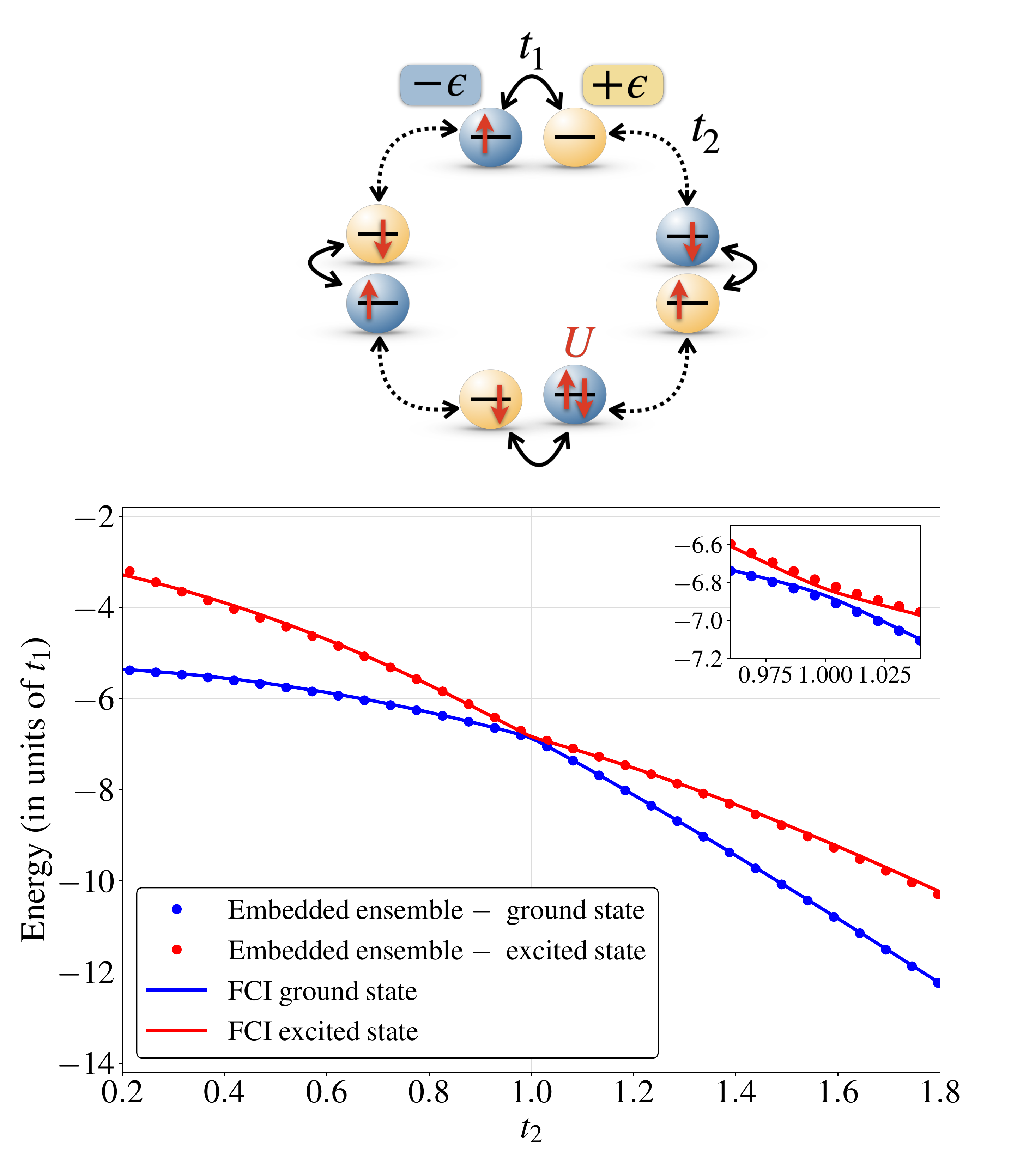}
\caption{\textbf{Top panel:} A schematic picture of the Hubbard ring model system. \textbf{Bottom panel:} The FCI ground and first excited singlet states (blue, and  red lines respectively), and embedding results for the ground and first excited state (blue and red dot markers, respectively) for the Hubbard ring model at $U=2$, $\epsilon = 1/2$, $t_1 = 1$ and different values of $t_2$.}
\label{fig:Fig4}  
\end{figure}

The first system on which we apply our embedding strategy is a Hubbard-like ring with an alternating hopping term and different local orbital energies on odd and even sites (see the top panel of Figure~\ref{fig:Fig4} for a sketch of the model). The model Hamiltonian for the ring reads as
\be\label{eq:Hubbard_ring_Hamiltonian}
\begin{aligned}
\hat{H} &= -\sum_{p=1}^{L}t_p\sum_{\sigma=\uparrow,\downarrow} 
(\cdop_{p,\sigma}\cop_{p+1,\sigma}+\cdop_{p+1,\sigma}\cop_{p,\sigma})
\\
&+U\sum_{p=1}^{L}\hat{n}_{p,\uparrow}\hat{n}_{p,\downarrow}
\:+\:\epsilon\sum_{p=1}^{L}(-1)^p\hat{n}_{p},  
\end{aligned} 
\ee
where
\be
t_p = \begin{cases}
  t_1 & \text{if } p \text{ is odd} \\
  t_2 & \text{if } p \text{ is even}
\end{cases}
\ee
are the hopping terms, $U$ is the local electronic repulsion, $\epsilon$ is the local orbital energy, $\hat{n}_{p,\sigma}=\cdop_{p,\sigma}\cop_{p,\sigma}$ is the occupation number operator for spin $\sigma$ on site $p$, and $\hat{n}_{p}=\sum_{\sigma=\uparrow,\downarrow}\hat{n}_{p,\sigma}$ is the total
occupation number operator on site $p$. In our FCI and embedding calculations, we chose the model with $L=8$ sites and $N=8$ electrons in total. To enforce the ring geometry, periodic boundary conditions ($\cdop_{L+1,\sigma} = \cdop_{1,\sigma}$) were imposed
in Eq.~\eqref{eq:Hubbard_ring_Hamiltonian}. The ratio $t_2 / t_1$, which controls the 
electron delocalization between the four ``dimers'' in the ring, is the key
parameter of study in this model. The $t_2 / t_1 \ll 1$ regime represents the bond breaking scenario, where the molecule turns into four separate dimers, while $t_2 / t_1 \gg 1$ represents a molecule with strong overlap between dimers. An interesting feature
of this model is a very close avoided crossing between the ground and first excited states occurring at $t_2 / t_1 = 1$ for $U=2t_1$ and $\epsilon = 0.5 t_1$, where the ground
and first excited state energies differ as
$E_1-E_0\approx3.55\times10^{-2}t_1$ (see also the inset on Figure~\ref{fig:Fig4}). Indeed,
around an avoided crossing, the eigenstates
of a non-interacting system usually undergo non-trivial mixing when electron interactions are switched on, which makes the present model a perfect
test example for our method.

Turning to the discussion of embedding
calculations, in the first step
of the process, we solve the noninteracting electron problem, which is, in the present case, the
tight-binding ring ($U=0$ in Eq.~\eqref{eq:Hubbard_ring_Hamiltonian}).
From the molecular orbitals of
the tight-binding ring, we then build
the equiensemble 1-RDM (where $w_0 = w_1 = 1/2$) of the ground and first excited noninteracting states in the lattice site basis, which
is a convenient and commonly used orthonormal basis
for the extraction of bath orbitals
in lattice models. As explained in
Section~\ref{sec:ensemble_embedding_two-state}, the application of three
successive Householder
transformations on the 1-RDM
in the lattice site basis produces
a disentangled cluster with
three bath orbitals connected to the fragment, which are used to build
the correlated clusters
in subsequent embedding calculations.
After obtaining the bath orbitals,
we perform single-shot embedding
calculations with the strategy described in Section~\ref{sec:Applications_general_description}.
The bottom panel of
Figure~\ref{fig:Fig4}
displays the embedding and FCI energies for the ground and first excited states of the
interacting ($U=2 t_1$) ring, for different
values of $t_2$ (at $t_1=1$). We observe that embedding results are in very good agreement with FCI energies for both states, especially in the region of the avoided crossing. Surprisingly, we also observe that there is no chemical potential tuning necessary to optimize the global number of electrons as long as the mean-field description of electron interactions between the 
correlated embedding cluster and cluster's environment $\mykettight{\Phi_0^{{E}}}$ are properly included in embedding calculations (interestingly, the same observation was reported by Marécat \textit{et al.} in the application of their self-consistent embedding scheme on the 1D Hubbard model~\cite{marecat2023unitary}). 
Moreover, we also report that any variations of ensemble
weights in the range $0<(w_0,w_1)<1$ (keeping in mind that they sum up to $1$) had no discernible differences
in the final evaluation of energies
from embedding clusters.

\subsection{System of hydrogen atoms}

\begin{figure}[!t]
\includegraphics[width=0.53\textwidth]{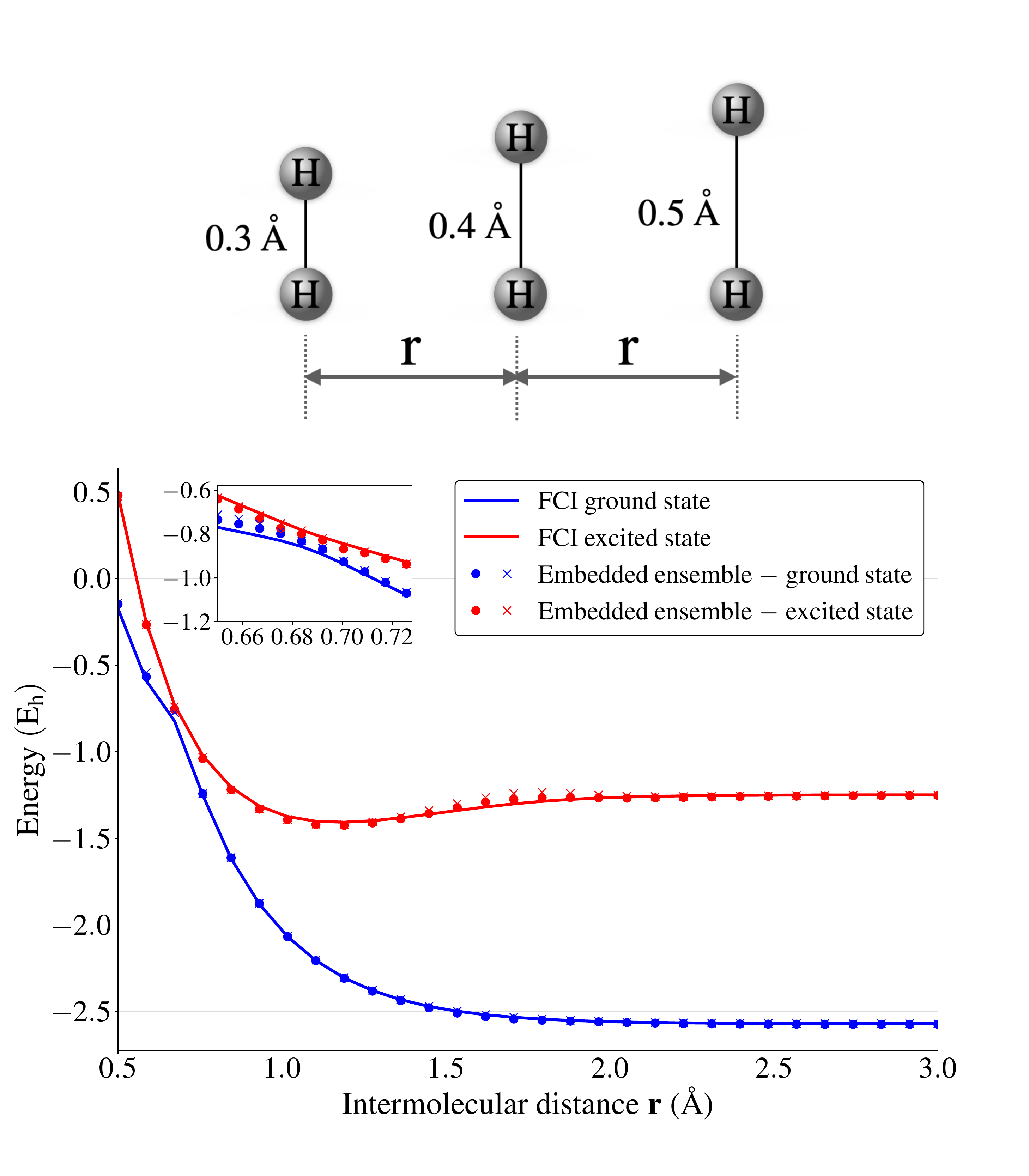}
\caption{\textbf{Top panel:} A schematic picture of the system of Hydrogen atoms by Tran \textit{et al.}~\cite{tran2019using}. \textbf{Bottom:} Dissociation curves of the FCI ground and first excited singlet states (blue and red lines, respectively), and embedding results for the ground and first excited state (blue and red markers, respectively) for the system of Hydrogen atoms. The embedding results are plotted with and without chemical potential optimization (dot ($\bullet$) and cross ($\times$) markers, respectively).}
\label{fig:Fig5}  
\end{figure}

For our second example,
we have chosen an \textit{ab initio} system
comprising an arrangement of six hydrogen atoms, which was originally used by Tran \textit{et al.} in their work on extending DMET to excited states~\cite{tran2019using}.
Similarly to what we had before in the Hubbard ring model,
this system also exhibits
an avoided crossing, making it an appealing choice
for a complementary study to a model system.
For an image
of the \textit{ab initio} system, see the top panel
of Figure~\ref{fig:Fig5}. 

The starting mean-field description of the full system was obtained from 
Restricted Hartree-Fock (RHF) calculations, from which we extracted the molecular orbitals for building the ensemble 1-RDM. The self-consistent
iterations of RHF scheme,
together with the calculation of one- and two-electron
integrals, were carried out with Psi4 python package~\cite{smith2020psi4}, using the STO-3G orbital basis set. 
\rev{Near the avoided crossing,
roughly corresponding to the intermolecular distances between $0.69$ and $0.72$, the use of second-order convergence method (SOSCF in Psi4) was employed for enforcing the convergence of RHF iterations.} Unlike in previous example,
the construction of a local and orthonormal orbital basis had to be carried
out before one could delineate fragments for embedding calculations.
This additional preprocessing step of orbital localization is
necessary for all \textit{ab initio} systems, with different
procedures employed in practice (for instance, see Ref.~\cite{koridon2021orbital} and reference therein).
For this system, we employed L\"{o}wdin's method
to build the symmetrically \textit{Orthogonalized Atomic Orbitals} (OAOs)~\cite{lowdin1950non,carlson1957orthogonalization,mayer2002lowdin}.
The OAOs are simple to obtain, and are
designed to resemble as closely as possible the
atomic orbitals of a chemical system, in our case
the $1s$ orbitals around Hydrogen atoms. 
After localization, the mean-field ensemble 1-RDM
was constructed, where we again chose
the equiensemble weight values.
Then, we ran embedding calculations
for different intermolecular distances (denoted by $\mathbf{r}$),
studying the dissociation of the hydrogen arrangement
into three separate molecules.
Concerning
the effect of the variation of ensemble weights,
we report the same observations as in
the Hubbard ring.
In contrast to the ring,
we observe that embedding calculations
in the system of hydrogen atoms
are amenable to the change in chemical potential.
However, as shown in the bottom panel of Figure~\ref{fig:Fig5}, the embedding results
are in excellent agreement with FCI
values for both ground
and excited states, even in the region near avoided crossing (in between $\mathbf{r}\approx0.66$\AA\: and $\mathbf{r}\approx0.72$\AA, see the inset in Figure~\ref{fig:Fig5}),
and the effect of chemical potential optimization
on the estimated energies is almost
inconsequential.
As a final point, in comparison to the state-specific
approach of Tran \textit{et al} (see Figure 2 in Ref.~\cite{tran2019using}), our strategy produced just as good results for all values of $\mathbf{r}$ considered.
While in their case, bath orbitals for each individual
state are computed from separate SCF calculations, in our strategy we rely
on the bath orbitals from a single SCF calculation for embedding
all states simultaneously which seems to be less costly.

\section{Conclusion and perspectives}\label{sec:Conclusion}

In the present paper we have tackled
the challenge of 
describing electronic excited
states using the methodology of
quantum embedding.
While the majority of previous
studies exploring
the feasibility of embedding
excited states have focused on state-specific strategies, we embarked
in this work in the direction
of an embedding
strategy for multiple (\textit{i.e.} ground and excited) states, guided by the recent
developments in DMET.

In the first part of the paper,
we touched upon the theoretical
aspects of fragment embedding,
beginning with a short review of
DMET and the use of
Householder transformation as the
key ingredient
in the clusterization 
of the ground-state 1-RDM,
enabling the
optimal orbital-space
partitioning at the mean-field
level of electronic
structure description. 
While the idempotency property of
1-RDMs, which is of key importance in ground-state DMET,
is lost when moving to ensembles of states, 
we have shown that an
exact clusterization for ensembles
is still achievable through the
application of a finite number of successive Householder transformations
on ensemble 1-RDMs. In the particular
case of the two-state ensemble consisting of the
linear combination of ground and
singly-excited mean-field states,
we have shown that the application 
of three Householder transformations
strictly block-diagonalizes
the ensemble 1-RDM, giving rise
to an enlarged, but still decoupled
cluster, containing a finite number
of bath orbitals and an integer number of electrons.
We also inquired into the possibility
of embedding larger-sized ensembles,
which we supported by numerical investigations. 
The latter have shown
that a decoupled cluster
can also be obtained in more involved
scenarios. Precisely,
we demonstrated that the Householder cluster size
scales linearly with the
size of the fractionally-occupied
natural orbital subspace in the
ensemble 1-RDM, while the number
of electrons inside the cluster
scales linearly with the number
of electrons, inhabiting these
fractionally-occupied natural orbitals. These results were
also corroborated by rigorous mathematical proofs in the single-orbital fragment case. A rationalization and a formal extension of the approach to the multi-orbital case have finally been obtained by extending the conventional DMET bath construction to GOK ensembles.

In the second part of this work,
we presented an implementation
of a single-shot embedding strategy
for description of the energies
of ground and first-excited states
in finite systems. Test calculations
of energy curves on two illustrative
examples have demonstrated that
embedding multiple states is a
very relevant strategy, especially
when different states become
increasingly closer in energy 
and as such are best treated in
an unbiased manner. The very strong
agreement of embedding and FCI
results in the case of system
of hydrogen atoms is also very promising, and tempts further 
exploration of the ensemble
embedding in other quantum chemistry
applications, like, for example,
describing conical intersections,
and dissociations
of heteroatomic molecules.

In summary, we demonstrated the
feasibility of embedding multiple
states by repurposing
the Householder transformation
in a straightforward manner
to generate additional bath 
orbitals in generally
non-idempotent ensemble 1-RDMs.
The variety of
insights, produced in this work
begets new questions and considerations
for improvements.
Firstly, let us mention
the prospect of extending
the single-shot embedding
strategy to more than two states.
In this study we
relied on extrapolating the
exact ensemble
emedding with a single
HOMO-LUMO excitation in the noninteracting limit,
to target the ground and first excited states of interacting systems.
Although the aforementioned numerical investigations
and the comparison with DMET
suggested that embedding ensembles with more excitations
is plausible, whether
the present strategy would
by analogy
produce reasonable results
for the second and higher excited states is, though
fascinating to contemplate, a difficult question
which we intend to address in future work.
In a different direction,
making the embedding self-consistent
would be one of the next logical
steps. As mentioned in
Section~\ref{eq:comparison_with_DMET},
combining many-electron ensembles
with local potential
functional embedding theory~\cite{sekaran2022local,yalouz2022quantum} would be an intriguing route to explore. In the perspective of ensemble density functional theory~\cite{Cernatic2022}, the embedding would become in-principle exact as both the embedding cluster and the reference full-size noninteracting (Kohn--Sham) system would reproduce the exact ensemble density (i.e., the ensemble occupation of the localized orbitals in the present context).    
Additionally, we stress
that the current implementation
was developed only for single-orbital fragments. 
It would
be interesting to investigate whether
the application of block-Householder
transformations~\cite{AML99_Rotella_Block_Householder_transf,sekaran2021householder,yalouz2022quantum,sekaran2023unified} can be re-used for
multi-orbital fragment embedding in
ensembles. Unlike in the standard DMET bath construction (or its extension to ensembles, as also discussed in this work), such transformations would automatically provide a complete orthonormal basis of the full one-electron Hilbert space (i.e., for the embedding cluster and its environment). That basis could be exploited in post-DMET treatments for recovering correlations that are external to the cluster, for example~\cite{sekaran2021householder}. 
As a final remark, excited states
are not the only domain
where the problem of non-idempotent
density matrices occurs.
\rev{Dealing with non-idempotency
is also unavoidable in ground-state
embedding with
correlated bath orbitals, where the approximate starting
description of the system
is no longer a single Slater determinant~\cite{sekaran2023unified}, and in finite-temperature DMET~\cite{sun2020finite}, where
the density matrix is non-idempotent due to temperature-dependent fractional orbital occupations. While
finite-temperature DMET starts with
nonzero fractional occupations in {\it all} orbitals
of the single-electron Hamiltonian,
which is a natural initial
description of extended systems
with continuous energy spectra,
the (GOK~\cite{gross1988rayleigh}) ensemble embedding philosophy
described in the present work deals with density matrices having
only a small number of fractionally
occupied orbitals. This is the primary reason why it can be made exact (for a non-interacting or mean-field ensemble) with a bath of limited size, unlike in finite-temperature DMET~\cite{sun2020finite}. For this
reason also, we strongly believe
that the computational strategy elaborated here
is better suited
for excited-state embedding
in finite systems. In addition,
it could also be straightforwardly
implemented in ground-state embedding
with correlated bath orbitals,
where the system is described
by an ensemble of states
in the natural
orbital basis. Work is in progress in these directions.}

\section*{Acknowledgements}
The authors thank ANR (CoLab project, grant no.: ANR-19-CE07-0024-02) for funding.

\section*{Conflict of interest}
The authors declare that they have no conflict of interest.

\section*{Data availability}
The data that support the findings of this study are available from the corresponding author upon reasonable request.

\appendix\label{sec:Appendix}

\section{Mathematical details of the tri-diagonalization by Householder transformations}\label{appendix:clus_succ_householder}

Let $\bgam$ be a symmetric matrix
with dimension $L$. Then, for every ${1\leq i \leq L-2}$, we consider the
following change of basis,
\be
\bgamt^{(i+1)} = {\bQ^{(i)\,\dagger}}\bgam\bQ^{(i)},
\ee
where
\be\label{eq:Q_transform_def}
\bQ^{(i)} = \bP^{(1)}\bP^{(2)}\dots\bP^{(i)},
\ee
and $\bP^{(i)}$ are
individual Householder transformations.
The latter are defined for $1\leq i \leq L-2$ as follows,
\be\label{eq:HHtr_def_i}
\bP^{(i)} &= \mathbf{I} - 2\mathbf{v}^{(i)}\mathbf{v}^{(i)\,\dagger}
\ee
where $\mathbf{I}$ is the $L\times L$ identity matrix, and $\mathbf{v}^{(i)}$ is the $i$-th Householder vector, whose elements are given by
\be
\mathbf{v}^{(i)} &= 
\begin{bmatrix}
v^{(i)}_1 \\
v^{(i)}_2 \\
\vdots \\
v^{(i)}_{L} \\
\end{bmatrix},
\ee
where
\begin{equation}\label{eq:Success_HH_eqs_1}
    \begin{aligned}
        &v_{1\leq j \leq i}^{(i)} = 0,
        \\
        &v_{i+1}^{(i)} = \dfrac{\tilde\gamma_{i+1,i}^{(i)}-\xi^{(i)}}{2r^{(i)}},
        \\
        &v_{i+2\leq j \leq L}^{(i)} = \dfrac{\tilde\gamma_{j,i}^{(i)}}{2r^{(i)}},
    \end{aligned}
    \\
\end{equation}
and
\begin{equation}\label{eq:Success_HH_eqs_2}
    \begin{aligned}
        &\xi^{(i)}=-sgn\left(\tilde\gamma_{i+1,i}^{(i)}\right)
        \sqrt{\sum_{j=i+1}^{L}\left(\tilde\gamma_{j,i}^{(i)}\right)^2},
        \\
        & r^{(i)} = \sqrt{\dfrac{1}{2}\left(\left(\xi^{(i)}\right)^2-
        \tilde\gamma_{i+1,i}^{(i)}\xi^{(i)}\right)}.
    \end{aligned}
\end{equation}
Starting with $\bgamt^{(1)}=\bgam$, the formulae in Eqs.~\eqref{eq:Success_HH_eqs_1} and~\eqref{eq:Success_HH_eqs_2} can be used to systematically construct, for every step $i$, the Householder transformation $\bP^{(i)}$
from the matrix elements of $\bgamt^{(i)}$, until reaching the fully tri-diagonal matrix $\bgamt^{(L-1)}$.
Given these definitions, one can show that the product of successive Householder transformations $\bQ^{(i)}$, is unitary.
This follows from the fact
that each Householder vector
$\mathbf{v}^{(i)}$
is normalized (see Eqs.~\eqref{eq:Success_HH_eqs_1} and~\eqref{eq:Success_HH_eqs_2}),
\be\label{eq:HHtr_vector_normalized}
\begin{aligned}
\mathbf{v}^{(i)\,\dagger}\mathbf{v}^{(i)} &=
\dfrac{(\gamt_{i+1,i}^{(i)}-\xi^{(i)})^2 - \sum_{j>i+1}^{L}(\gamt_{j,i}^{(i)})^2}{4(r^{(i)})^2}
\\
&=\dfrac{\sum_{j=i+1}^{L}(\gamt_{j,i}^{(i)})^2 - 2\gamt_{i+1,i}^{(i)}\xi^{(i)} + (\xi^{(i)})^2}{4(r^{(i)})^2}
\\
&=\dfrac{2\left((\xi^{(i)})^2 - \gamma_{i+1,i}^{(i)}\xi^{(i)}\right)}{4(r^{(i)})^2} = 
\dfrac{{4(r^{(i)})^2}}{{4(r^{(i)})^2}} = 1,
\end{aligned}
\ee
and hence, according to Eqs.~\eqref{eq:HHtr_def_i},
and~\eqref{eq:HHtr_vector_normalized},
every Householder transformation $\bP^{(i)}$ is unitary
and Hermitian,
\be
\begin{aligned}
\bP^{(i)\,\dagger}\bP^{(i)} &= \bP^{(i)}\bP^{(i)\,\dagger} = \bP^{(i)}\bP^{(i)}
\\
&= \left( \mathbf{I} - 2\mathbf{v}^{(i)}\mathbf{v}^{(i)\,\dagger} \right)
\left( \mathbf{I} - 2\mathbf{v}^{(i)}\mathbf{v}^{(i)\,\dagger} \right)
\\
&=
\mathbf{I} - 4\mathbf{v}^{(i)}\mathbf{v}^{(i)\,\dagger}
+ 4\mathbf{v}^{(i)}\left(\mathbf{v}^{(i)\,\dagger}\mathbf{v}^{(i)}\right)\mathbf{v}^{(i)\,\dagger}
\\
&=\mathbf{I} - 4\mathbf{v}^{(i)}\mathbf{v}^{(i)\,\dagger}
+ 4\mathbf{v}^{(i)}\mathbf{v}^{(i)\,\dagger}
\\
&=\mathbf{I}.
\end{aligned}
\ee
It is then trivial to show that $\bQ^{(i)\,\dagger}\bQ^{(i)}=\bQ^{(i)}\bQ^{(i)\,\dagger}=
\mathbf{I}$.
For a complete derivation of the above formulae and the \textit{tri-diagonalization} procedure, the reader is referred to the book Numerical Analysis, 9th edition by Burden and Faires~\cite{burden2011numerical}.

\section{Technical details on the test calculations
for exact embedding 
of larger ensembles} 
 \label{app:num_method}


For the production of numerical data in
Fig.~\ref{fig:Fig3} in Section~\ref{sec:exact_embedding_NI_ensembles} we used
a uniform and non-interacting
Hubbard chain model with $20$
lattice sites. The Hamiltonian
for this model
can be obtained from Eq.~\eqref{eq:Hubbard_ring_Hamiltonian} by setting $t_1=t_2=1$, $U=0$
and $\epsilon=0$, and $\cdop_{L+1,\sigma}=0$ to ensure open boundary conditions for the chain geometry.

In order to unravel
in exact embedding
the behavior of
the properties
of the Householder
cluster with increasing 
the size of the ensemble,
we construct an artificial
ensemble 1-RDM, containing
a fractionally occupied
natural orbital subspace
of modifiable dimensionality.
In the lattice site basis,
the ensemble 1-RDM is built
as follows,
\be\label{eq:1RDM_artificial}
\gamma^{\bw}_{pq} = \sum_{i=1}^L f_i C_{p i} C_{q i}
\ee
where $\{C_{p i}\}$ are the
molecular orbital coefficients
of the noninteracting Hubbard
chain, and
\be\label{eq:1RDM_artificial_fractional_sector}
f_i = \begin{cases}
    1,&  1 \leq i\leq \frac{L}{4}\\
    \frac{N_{frac}^{elec}+\delta N_{frac}^{orb}(N_{frac}^{orb}-2i+\frac{L}{2}+1)}{2N_{frac}^{orb}}, &
    \frac{L}{4} < i \leq \frac{L}{4}+N_{frac}^{orb} \\
    0,& \frac{L}{4}+N_{frac}^{orb} < i \leq L,
\end{cases}
\ee
are the orbital occupation numbers.
The $L/4$ orbitals
with the lowest orbital energies
are fully occupied. The
next $N_{frac}^{orb}$ orbitals higher
in energy are artificially
constrained to fractional
occupations
(with ${0<f_i<1}$).
In the latter sector,
$f_i$ are strictly decreasing
with $i$, with the rate
depending on the parameter
$\delta$ (set to $2.5\times10^{-2}$ in
the present work), and
sum up
to $N_{frac}^{elec}/2$ electrons.
The remaining ${3L}/{4}-N_{frac}^{orb}$ orbitals
are left unoccupied.
By construction, $f_i$
in Eq.~\eqref{eq:1RDM_artificial_fractional_sector} sum up to
\be
\sum_{i=1}^L f_i = \frac{L}{4} + \dfrac{N_{frac}^{elec}}{2}.
\ee
For obtanining the results, displayed in Fig.~\ref{fig:Fig3},
we applied successive
Householder transformations
as described in Appendix~\ref{appendix:clus_succ_householder} on
the 1RDM in
Eq.~\eqref{eq:1RDM_artificial},
until reaching a block-diagonal structure. In addition, as already
explained in Section~\ref{sec:exact_embedding_NI_ensembles}, we removed
from Fig.~\ref{fig:Fig3} all data points where  $N_{frac}^{elec}\rev{\geq} 2N_{frac}^{orb}$, since they
correspond to 1-RDMs with unphysical fractional orbital occupations (\textit{i.e.,} $f_i\geq 1$).

\section{A rigorous proof of the properties of
Householder clusters in non-idempotent 1RDMs}\label{appendix:frac_rdm_general_math_proofs}

\subsection{A general proof}

In the following, we demonstrate
for a given 1-RDM in the Householder basis,
that under certain conditions
on its eigenvalues, 1) the Householder cluster and Householder environment are strictly disentangled, and 2) the trace of the Householder cluster is given by the sum over distinct eigenvalues of the 1-RDM.

Let us denote by $\bgam$ the 1-RDM in the local basis. Suppose we perform the change of basis $\bgamt=\bQ^{\dagger}\bgam\bQ$, where
$\bQ$ is a product of $m-1$ Householder transformations (see Appendix~\ref{appendix:clus_succ_householder}),
and $m<L$. Then, in the structure of
$\bgamt$ we can identify three different sub-blocks,
\be\label{eq:1RDM_after_m_householders}
\bgamt =
\begin{bmatrix}
\bgamt_{CC} & \bgamt_{EC}^{\dagger} \\
\bgamt_{EC} & \bgamt_{EE}
\end{bmatrix},
\ee
where $\bgamt_{CC}$,
is the Householder cluster with $\dim(\bgamt_{CC})=m$, $\bgamt_{EE}$ is the cluster's environment
with $\dim(\bgamt_{EE})=L-m$,
and $\bgamt_{EC}$ is the environment-cluster
coupling sub-block with dimensions ${(L-m) \times m}$.
In general, $\bgamt_{CC}$ is a tridiagonal matrix, 

\be\label{eq:1RDM_m_householders_cluster}
\bgamt_{CC} =
\begin{bmatrix}
\gamt_{11}   & \gamt_{21}  & 0  & \dots &  0    \\[0.3em] 
\gamt_{21}  &  \gamt_{22}  & \ddots  & \ddots   &  \vdots  \\[0.3em]
0 &\ddots&\;\;\ddots&&0    \\[0.3em]
\vdots  &  \;\ddots  & & &  \gamt_{m\,(m-1)}    \\[0.3em]
0 & \dots & 0  & \;\;\gamt_{m\,(m-1)}  &  \gamt_{mm}   \\[0.3em]
\end{bmatrix},
\ee
where $[\bgamt_{CC}]_{ij}\overset{|i-j|>1}{=}0$, and
$\bgamt_{EC}$ has a single nonzero column, \textit{i.e.,}
\be\label{eq:1RDM_m_householders_env_cluster}
\bgamt_{EC} =
\begin{bmatrix}
0 &  \dots   & 0  &  \gamt_{(m+1)\,m}    \\[0.3em]
0 &  \dots   & 0  &  \gamt_{(m+2)\,m}    \\[0.3em]
\vdots &    & \vdots  &  \vdots    \\[0.3em]
0 &  \dots   & 0  &  \gamt_{L\,m}    \\[0.3em]
\end{bmatrix}.
\ee
If $\bgam$ has
exactly $m$
{\it distinct} eigenvalues $\{f_k\}_{k}^m$ (natural orbital occupations
in the eigen-orbital basis),
it fulfills, according
to the Cayley-Hamilton theorem,
the following minimal polynomial
equation of degree $m$,
\be\label{eq:gamma_minimal_polynomial_expanded}
\bgam^m - \left(\sum_{k}^m f_k\right)\bgam^{m-1}
+ \dots + (-1)^m \left(\prod_{k}^m f_k\right)\mathbf{I} = \bm{0}.
\ee
By invoking the above equation for $\bgamt$, 
it is possible
to show that $\bgamt_{CC}$ is
decoupled from $\bgamt_{EE}$,
by demonstrating that
\be\label{eq:1RDM_general_cluster_decoupling}
\bgamt_{EC}=\bm{0}_{EC}.
\ee
In that particular case, it
also follows that $\bgamt_{CC}$ contains
an easily calculable number of electrons, given by the summation
over distinct occupation numbers,
\be\label{eq:1RDM_general_cluster_number_of_electrons}
\Tr[\bgamt_{CC}]=\sum_{k}^m f_k.
\ee
In order to prove that the cluster
is decoupled, we first prove by induction
the following statements for
the powers of $\bgamt$. For any $2\leq l\leq m-1$ ($2\leq l\leq m$, respectively),
\begin{subequations}\label{eq:decoupling_lemma}
\begin{align}
[\bgamt^l]_{ij} &\overset{i>m,\;j<m-l+1}{=}0,\label{eq:decoupling_lemma_eq1}
\\
[\bgamt^l]_{ij} &\overset{i>m,\;j=m-l+1}{=}\gamt_{im}
\gamt_{m\,(m-1)}\dots\gamt_{(m-l+2)\,(m-l+1)}.\label{eq:decoupling_lemma_eq2}
\end{align}
\end{subequations}
For the base case $l=2$, we have,
\be\label{eq:1RDM_power_2_ij}
[\bgamt^2]_{ij}=\sum_{k=1}^L \gamt_{ik}\gamt_{kj}=
\sum_{k=1}^{m-1} \gamt_{ik}\gamt_{kj}
+ \sum_{k=m}^L \gamt_{ik}\gamt_{kj}.
\ee
According to Eq.~\eqref{eq:1RDM_m_householders_env_cluster},
\be
\gamt_{ik}\overset{i>m,\;k\leq m-1}{=}0,
\ee
and
\be
\gamt_{kj}\overset{k\geq m,\;j<m-1}{=}0.
\ee
Therefore, for $j<m-1$,
and $j=m-1$, respectively, Eq.~\eqref{eq:1RDM_power_2_ij}
reads as
\begin{subequations}
\begin{align}
[\bgamt^2]_{ij} &\overset{i>m,\;j<m-1}{=}0,
\\
[\bgamt^2]_{ij} &\overset{i>m,\;j=m-1}{=}\gamt_{im}\gamt_{m\,(m-1)}.    
\end{align}
\end{subequations}
For $l\geq2$, we assume that Eq.~\eqref{eq:decoupling_lemma}
holds for some $l$, and examine the case $l+1$,
\be\label{eq:1RDM_power_lplus1_ij}
\begin{aligned}
[\bgamt^{l+1}]_{ij}&=\sum_{k=1}^L [\bgamt^{l}]_{ik}\gamt_{kj}
\\
&= \sum_{k=1}^{m-l}[\bgamt^{l}]_{ik}\gamt_{kj} +
\sum_{k=m-l+1}^{L}[\bgamt^{l}]_{ik}\gamt_{kj}.  
\end{aligned}
\ee
In the above equation, we look for matrix elements with $i>m$ and $j\leq m-l$. According to our assumption for the case $l$,
\be\label{eq:1RDM_power_l_ik}
[\bgamt^{l}]_{ik} \overset{i>m,\;k\leq m-l}{=}0,
\ee
and, according to Eq.~\eqref{eq:1RDM_m_householders_cluster},
\be\label{eq:1RDM_kj}
\gamt_{k\,j}\overset{m-l<k\leq L,\;j<m-l}{=}0.
\ee
For $j<m-l$, we conclude from Eqs.~\eqref{eq:1RDM_power_lplus1_ij}~-~\eqref{eq:1RDM_kj} that
\be
[\bgamt^{l+1}]_{ij}\overset{i>m\;,j<m-l}{=}0,
\ee
while for $j=m-l$, we conclude from Eq.~\eqref{eq:1RDM_power_lplus1_ij} that
\be
\begin{aligned}
[\bgamt^{l+1}]_{ij} &\overset{i>m,\;j=m-l}{=}
[\bgamt^{l}]_{i\,(m-l+1)}\gamt_{(m-l+1)\,(m-l)}
\\
&\quad\;\;\:=\gamt_{im}\gamt_{m\,(m-1)}\dots\gamt_{(m-l+1)\,(m-l)}.    
\end{aligned}
\ee
This completes the proof of Eq.~\eqref{eq:decoupling_lemma}.
As a side note, Eq.~\eqref{eq:decoupling_lemma_eq1}
no longer holds for $l=m$. Otherwise,
the index $j$ would attain the values
$j<m-m+1=1$, which is absurd since $j\geq1$.

Now we are ready to prove Eq.~\eqref{eq:1RDM_general_cluster_decoupling}.
The key step is to evaluate the minimal polynomial of $\bgamt$ (see Eq.~\eqref{eq:gamma_minimal_polynomial_expanded}) for elements with $i>m$ and $j=1$,
\be
\begin{aligned}
[\bgamt^{m}]_{i1}&-\left(\sum_{k}^m f_k\right)
[\bgamt^{m-1}]_{i1}+\dots
\\
&+(-1)^m \left(\prod_{k}^m f_k\right)[\mathbf{I}]_{i1} = 0.
\end{aligned}
\ee
According to Eq.~\eqref{eq:decoupling_lemma_eq1},
all terms $[\bgamt^{l}]_{i1}$ where $l<m$, vanish. Therefore, we are left with,
\be
[\bgamt^{m}]_{i1} \overset{i>m}{=} 0,
\ee
which implies (see Eq.~\eqref{eq:decoupling_lemma_eq2}),
\be\label{eq:decoupling_gamma_im_product}
\gamt_{im}\gamt_{m\,(m-1)}\dots\gamt_{32}\gamt_{21} \overset{i>m}{=} 0.
\ee
Since $\gamt_{m\,(m-1)},\dots,\gamt_{32},\gamt_{21}\neq0$ (we
assume the first $m-1$ Householder transformations
are defined), we are left to conclude
that
\be\label{eq:env-cluster-coupling-zero}
\gamt_{im}\overset{i>m}{=}0 \;\;\Leftrightarrow \;\; \bgamt_{EC}=\bm{0}_{EC}.
\ee
Consequently, $\bgamt$ (see Eq.~\eqref{eq:1RDM_after_m_householders}) has 
the expected block-diagonal structure,
\be
\bgamt =
\begin{bmatrix}
\bgamt_{CC} & \bm{0}_{EC}^{\dagger} \\
\bm{0}_{EC} & \bgamt_{EE}
\end{bmatrix}.
\ee

Next, we prove the cluster trace formula in Eq.~\eqref{eq:1RDM_general_cluster_number_of_electrons}, by making use of the following statements for $\bgamt_{CC}$, which we prove by induction. For any $2\leq l \leq m$,
\be\label{eq:gamma_CC_lemma1}
[\bgamt_{CC}^l]_{l1} = \left(\sum_{k=1}^l \gamt_{kk}\right)[\bgamt_{CC}^{l-1}]_{l1},
\ee
for any $2\leq l \leq m-1$,
\be\label{eq:gamma_CC_lemma2}
[\bgamt_{CC}^l]_{(l+1)\,1} = \gamt_{(l+1)\,l}[\bgamt_{CC}^{l-1}]_{l1},
\ee
and any $2\leq l \leq m-2$,
\be\label{eq:gamma_CC_lemma3}
[\bgamt_{CC}^l]_{i\,1} \overset{i>l+1}{=}0.
\ee
The case $l=2$ is readily verified
using Eq.~\eqref{eq:1RDM_m_householders_cluster},
\be
[\bgamt_{CC}^2]_{21} = \sum_{k=1}^m \gamt_{2k}\gamt_{k1} = \left(\gamt_{11}+\gamt_{22}\right)\gamt_{21},
\ee
\be
[\bgamt_{CC}^2]_{31} = \sum_{k=1}^m \gamt_{3k}\gamt_{k1} = \gamt_{32}\gamt_{21},
\ee
and
\be
[\bgamt_{CC}^2]_{i1} \overset{i>3}{=} \sum_{k=1}^m \gamt_{ik}\gamt_{k1} = 0.
\ee
Now let us assume that Eqs.~\eqref{eq:gamma_CC_lemma1}~-~\eqref{eq:gamma_CC_lemma3} hold for a given $l\geq2$,
and verify for $l+1$.
We start with Eq.~\eqref{eq:gamma_CC_lemma3},
\be
[\bgamt_{CC}^{l+1}]_{i1} 
= \sum_{k=1}^m \gamt_{ik}[\bgamt_{CC}^{l}]_{k1}
\overset{i>l+2}{=}
\sum_{k=l+2}^m \gamt_{ik}[\bgamt_{CC}^{l}]_{k1}.
\ee
Given the assumption for $l$, ${[\bgamt_{CC}^{l}]_{k1}\overset{k>l+1}{=}0}$, hence
\be
[\bgamt_{CC}^{l+1}]_{i1}\overset{i>l+2}{=}0.
\ee
For Eq.~\eqref{eq:gamma_CC_lemma2}, we have,
\be
[\bgamt_{CC}^{l+1}]_{(l+2)\,1} = 
\sum_{k=1}^m \gamt_{(l+2)\,k}[\bgamt_{CC}^{l}]_{k1}.
\ee
According to Eqs.~\eqref{eq:1RDM_m_householders_cluster} and~\eqref{eq:gamma_CC_lemma3} the only nonzero term is
\be
[\bgamt_{CC}^{l+1}]_{(l+2)\,1} = \gamt_{(l+2)\,(l+1)}[\bgamt_{CC}^{l}]_{(l+1)\,1}.
\ee
Finally, we prove Eq.~\eqref{eq:gamma_CC_lemma1}. For the case $l+1$, we have,
\be
[\bgamt_{CC}^{l+1}]_{(l+1)\,1} =
\sum_{k=1}^m \gamt_{(l+1)\,k} [\bgamt_{CC}^{l+1}]_{k\,1}.
\ee
According to Eqs.~\eqref{eq:1RDM_m_householders_cluster} and~\eqref{eq:gamma_CC_lemma3}, the only nonzero terms are
\be\hspace{-0.3cm}
[\bgamt_{CC}^{l+1}]_{(l+1)\,1} = \gamt_{(l+1)\,l} [\bgamt_{CC}^{l}]_{l\,1} + \gamt_{(l+1)\,(l+1)} [\bgamt_{CC}^{l}]_{(l+1)\,1}.
\ee
Inserting Eqs.~\eqref{eq:gamma_CC_lemma1} and~\eqref{eq:gamma_CC_lemma2} into the above equation, we obtain
\be
\begin{aligned}
[\bgamt_{CC}^{l+1}]_{(l+1)\,1} &= \gamt_{(l+1)\,l} \left(\sum_{k=1}^l \gamt_{kk}\right)[\bgamt_{CC}^{l-1}]_{l1}
\\
&+ \gamt_{(l+1)\,(l+1)} [\bgamt_{CC}^{l}]_{(l+1)\,1}
\\
&=\left(\sum_{k=1}^l \gamt_{kk}\right)[\bgamt_{CC}^{l}]_{(l+1)\,1} + \gamt_{(l+1)\,(l+1)} [\bgamt_{CC}^{l}]_{(l+1)\,1}
\\
&=\left(\sum_{k=1}^{l+1} \gamt_{kk}\right)[\bgamt_{CC}^{l}]_{(l+1)\,1}.
\end{aligned}
\ee
This completes the proof of Eqs.~\eqref{eq:gamma_CC_lemma1} --~\eqref{eq:gamma_CC_lemma3}. The proof of Eq.~\eqref{eq:1RDM_general_cluster_number_of_electrons}
now follows from Eq.~\eqref{eq:gamma_CC_lemma1} for $l=m$,
\be\label{eq:gamma_CC^m_m1}
[\bgamt^{m}_{CC}]_{m1} = \left(\sum_{i=1}^m \bgamt_{ii} \right)[\bgamt^{m-1}_{CC}]_{m1},
\ee
and Eq.~\eqref{eq:gamma_CC_lemma3} for $i=m$,
\be\label{eq:gamma_CC^l_m1}
[\bgamt^{l}_{CC}]_{m1} \overset{l<m-1}{=}0.
\ee
Inserting Eqs.~\eqref{eq:gamma_CC^m_m1} and
~\eqref{eq:gamma_CC^l_m1} into Eq.~\eqref{eq:gamma_minimal_polynomial_expanded},
it follows that
\be\label{eq:gamma_cluster_polynomial_zero_m-1}
\left[\left(\sum_{i=1}^m \bgamt_{ii} \right) -
\left(\sum_{k}^m f_k \right)\right]
[\bgamt^{m-1}_{CC}]_{m1} = 0.
\ee
Since $[\bgamt^{m-1}_{CC}]_{m1}\neq 0$, we conclude
from the above equation that
\be\label{eq:1RDM_cluster_trace_final}
\Tr[\bgamt_{CC}]\equiv\sum_{i=1}^m \bgamt_{ii} = 
\sum_{k}^m f_k.
\ee

In summary, the dimension
of the Householder cluster
and number of electrons inside the cluster
after reaching the decoupling will depend only
on the distinct occupation numbers of $\bgam$.
The application of successive Householder transformations
on the local basis
will extract, from each invariant subspace of $\bgam$ with occupation number $f_k$, a single bath orbital coupled to the fragment. 

\subsection{Particular examples}

\subsubsection{Two-state ensemble}
In the case of the two-state ensemble 1-RDM
of the noninteracting system
(see Sections~\ref{sec:ensemble_embedding_original_motivations} and~\ref{sec:ensemble_embedding_two-state}), there are four
different occupation numbers,
with ${f_1=\dots=f_{N/2-1}=1}$
(fully occupied orbitals) and
${f_{N/2+2}=\dots=f_{L}=0}$ (unoccupied orbitals). For the fractionally occupied HOMO and LUMO, the occupation
numbers read
$f_{N/2}=(2w_0 + w_1)/2$ and $f_{N/2+1}=w_1/2$ (see Eqs.~\eqref{eq:1-RDM_ensemble_MF} and~\eqref{eq:1-RDM_singly-excited}). Therefore,
\be
\dim(\bgamt^{\bw}_{{C}{C}}) = 4,
\ee
and
\be
\begin{aligned}
\Tr[\bgamt^{\bw}_{{C}{C}}] &= f_1 + f_{N/2} + f_{N/2+1} + f_{N/2+2}
\\
&= 1 + \dfrac{2w_0 + w_1}{2} + \frac{w_1}{2} + 0 = 2.  
\end{aligned}
\ee

\subsubsection{Larger ensembles}
In the numerical study at the end of Section~\ref{sec:ensemble_embedding_two-state} (see also Appendix~\ref{app:num_method}),
$\bgamt^{\bw}$ has $L/4$ fully occupied orbitals, $3L/4 - N_{frac}^{orb}$ unoccupied orbitals, and $N_{frac}^{orb}$ fractionally occupied orbitals with different occupations summing up to $N_{elec}^{frac}/2$.
In total,
there are $N_{frac}^{orb} + 2$
orbitals with different occupations, hence
\be
\dim(\bgamt_{CC}^{\bw}) = 2 + N_{frac}^{orb},
\ee
and, according to Eq.~\eqref{eq:1RDM_general_cluster_number_of_electrons},
\be
\Tr[\bgamt_{CC}^{\bw}] = 1 + {N_{elec}^{frac}}/{2}.
\ee

\subsubsection{Idempotent matrices}
Finally, as a special example,
let us investigate a generic matrix
with only two different occupation numbers, \textit{i.e.}
$f_1=1$ and $f_2=0$. According to Eq.~\eqref{eq:gamma_minimal_polynomial_expanded},
the matrix should fulfill the polynomial
\be
\bgam^2 - \left(0+1\right)\bgam = \bm{0},
\ee
which can be identified with the idempotency condition.
For any idempotent matrix,
it follows that a single Householder transformation $\bgamt=\bP^{(1)}\bgam\bP^{(1)}$
produces a cluster, decoupled from the environment. According
to Eq.~\eqref{eq:decoupling_gamma_im_product} and the
subsequent discussion,
\be
\gamt_{i2}\overset{i>2}{=}0.
\ee
From the idempotency, we also deduce,
using Eq.~\eqref{eq:1RDM_cluster_trace_final}, that the
cluster contains one electron per spin,
\be
\Tr[\bgamt_{CC}] = \gamt_{11} + \gamt_{22} = 1.
\ee
Thus, from the present analysis, we also recover the
properties of the embedding cluster in idempotent
matrices, previously derived in the context
of ground-state embedding (see Ref.~\cite{sekaran2021householder}).

\bibliography{biblio.bib}

\end{document}